# Development of a self-consistent thermodynamically optimized database along with phase transition experiments in Ni-Mn-Ga system for magnetocaloric applications

Nishant Tiwari[a], Varinder Pal[a], Swagat Das[a], and Manas Paliwal[*a]

[a]Department of Metallurgical and Materials Engineering, Indian Institute of Technology Kharagpur, 721302, West Bengal, India

Address to correspondence:

[*]manas.paliwal@metal.iitkgp.ac.in

**Abstract**

Magnetocaloric materials have received significant attention of research community as they can minimize the use of harmful gases (CFCs, HFCs) and render eco-friendly refrigeration. Heusler alloys ($Ni_2MnGa$) are known for their magnetocaloric effects, which make them useful as energy efficient and eco-friendly refrigerating materials. Magnetocaloric properties significantly depend on the composition of these alloys. Ni-Mn-Ga is one of the interesting Heusler systems, which exhibits magnetocaloric properties. In the present study, we performed the thermodynamic optimization of two sub binaries of the Ni-Mn-Ga system: Mn-Ga and Ni-Ga, using CALPHAD approach. A Modified Quasichemical Model (MQM) was used to describe the thermodynamic properties of the liquid solutions in both the binaries. Both the binaries were combined with Mn-Ni to develop a self-consistent thermodynamic database for Ni-Mn-Ga. In order to resolve the existing experimental discrepancies in the Mn-Ga and Ni-Ga system, few alloy compositions were prepared and analyzed using differential thermal analysis. Finally, the developed thermodynamic database was used to calculate the ternary isothermal section of the Ni-Mn-Ga (Heusler alloy) system at 1073 K with a proposed phase region for magnetocaloric applications.

## 1. Introduction

Technological advancements are key to human progress. These technologies need to be constantly upgraded or replaced with time to bear the new challenges. One of the technologies that has significantly improved the human life and facilitated us to work comfortably in every climate is atmospheric temperature control. This control can be achieved by using various cooling systems such as refrigeration and air conditioning. These cooling systems involve the usage of gases like chlorofluorocarbons (CFCs) and hydrofluoric carbon (HFCs). Release of these gases directly into the atmosphere depletes the ozone layer. This ozone layer is responsible for protecting earth against the harmful ultraviolent radiations. In addition, the rate of the ozone depletion is increasing constantly with continuous increase in the usage of these cooling devices. It is reported that cooling services are currently accounting for over 10 % of the greenhouse gas emissions which affects the ozone layer significantly[1]. One of the alternatives to encounter this problem is the use of magnetocaloric effect (MCE) for refrigeration. Magnetocaloric effect (MCE) is a characteristic of material, which can be initiated with the application of magnetic field. Like conventional refrigerators, a magnetic refrigerator uses magnetocaloric material to collect heat from a low temperature load (cold exchanger) and discharge it to a high temperature sink (hot exchanger). [3]. Thermodynamic relationships mentioned below provides a roadmap to understand the magnetocaloric materials.

$$\mathrm{RC}\ (\mathrm{H}) = \int_{\mathrm{T}_c}^{\mathrm{T}_h} \Delta S_m\left(\mathrm{T}, \mathrm{H}_{\text{applied}}\right) dT \quad (1)$$

$$\Delta S_m\left(\mathrm{T}, \mathrm{H}_{\text{applied}}\right) = \int_0^{H_{applied}} \left(\frac{\partial \mathrm{M}}{\partial \mathrm{T}}\right)_H dT \quad (2)$$

$$\Delta \mathrm{T}_{ad}\left(\mathrm{T}, \mathrm{H}_{\text{applied}}\right) = -\int_0^{H_{applied}} \frac{T}{C(T,H)} \left(\frac{\partial \mathrm{M}}{\partial \mathrm{T}}\right)_H dT \quad (3)$$

RC (refrigerant capacity), $\Delta S_m$, $T_h$, $T_c$, and $T_{ab}$ in equation 1 represent the amount of heat transferred between the reservoirs, total entropy change, hot reservoir temperature, cold reservoir temperature and the change in adiabatic temperature respectively. Similarly, in equation 2 and 3: $H_{applied}$, and M denotes applied magnetic moment, and magnetization. $C(T,H)$ in equation 3 denotes heat capacity as a function of magnetic field and temperature. All the variables mentioned in equations (1), (2) and (3) are important to determine the MCE of an MCM. As evident from equation 1, the refrigeration capacity is directly dependent on the entropy change of the material; hence, it is desired to maximize the entropy change of the material for its application as an MCM. Total entropy change of a material has three components: configurational (crystal lattice structure), non-configurational (atomic vibrations) and magnetic. For efficient refrigeration, the heat removed from the system is directly proportional to the total entropy change under isothermal conditions. Hence, it is imperative to maximize the configurational and magnetic contributions to the total entropy change in the material for achieving efficient refrigeration. Most importantly, the magnetic and configurational entropy change should be achieved around room temperature.

Magnetocaloric materials can be classified based on operating temperatures as: (1) low or intermediate (LMCM) and (2) room temperature magnetocaloric materials (RMCM). LMCM can be further classified into adiabatic demagnetization (ADR) and nuclear adiabatic demagnetization (NDR) materials. ADR and NDR are usually the paramagnetic materials, which cool on demagnetisation at low temperatures (5 to 1 K) [2]. $GdLiF_4$ and $Ce_2Mg_3.(N0_3)_{12}.24H_2O$ are reported ADR materials which shows MCE at as low as 0.6 mK temperature [3] while $^{19}F$ nuclei was reported to be the first NDR material with cooling from 1.2 K to 170 mK [4]. RMCM are MCM for cooling at room temperature with a favourable low heat capacity (see equation (3)), moderate thermal conductivity, low magneto thermal hysteresis etc. [5]. The phase transitions on the application of magnetic field to a MCM can magnify the MCE (increase the $\Delta S_m$ and $\Delta T_{ad}$ ). Whereas, in case of non-phase transition in MCM there is either small or moderate change in the $\Delta S_m$ and $\Delta T_{ad}$. RMCM are primarily Gd [6,7] and Gd-R (R-rare earth) based alloys [8,9], $Gd_5(Si,Ge)_4$ [10–12], $(Mn,Fe)_2(P,Si)$ [13],

La(Fe,Co,Mn,Si)$_{13}$Hy [14–16], Fe-Rh [17–20], Heusler alloys and other martensitic materials [21–24]. All the types of LMCM and RMCM have certain merits and demerits which are reported by Sandeman and Takei [2]. Heusler alloys are advantageous for MCM applications in comparison to other reported MCM alloys because of their ability to improve MCE by altering their composition. They exhibit martensitic (diffusion less) phase transition from a low temperature martensitic phase (disordered, $L1_0$) to high temperature austenitic phase (ordered, $L2_1$) [25,26]. $Ni_2MnGa$, $Ni_2MnSn$ and $Ni_2MnIn$ are the known Heusler alloys for MCE applications reported. Hu et al. [27] first time reported magnetocaloric properties of $Ni_{51.5}Mn_{22.7}Ga_{25.8}$ Heusler alloy. They observed positive entropy change ($\Delta S_m$~4 J $Kg^{-1}$ $K^{-1}$) for $Ni_{51.5}Mn_{22.7}Ga_{25.8}$ at 0.9 Tesla [27]. Entropy change can be enhanced by combining first order and the second order phase transitions. Initially for Heusler alloy, the first and second order phase transitions were found at different temperatures. Later, by changing the composition, researchers tried to merge both the phase transitions at same temperature. Enhancement in the entropy change ($\Delta S_m$~ -5 to -20 J $Kg^{-1}$ $K^{-1}$) by combining the first and second order transition of Heusler alloys was first reported by Pareti et al. [22].

As magnetic properties of MCMs can be altered easily by changing the composition, a critically assessed thermodynamic database of Ni-Mn-Ga based ternary alloy system and its sub binary systems can be helpful to tailor these properties. In the present study, thermodynamic assessment of Mn-Ga and Ni-Ga binary systems have been carried out. Detailed thermodynamic assessment of Mn-Ga binary system has not been reported till date. In addition, there are very limited studies on thermodynamic properties (enthalpy of formation, mixing enthalpy, activities) reported for the system. However, experimental studies showing the phase stabilities and transitions in Mn-Ga system are available in the literature. Experimental phase diagram using various thermal analysis and characterization techniques by different researchers are inconsistent with each other. As a result, there is significant variation in the reported phase diagrams of Mn-Ga binary system [28–30]. Recently, an evaluation of Mn-Ga system was carried out by Hao and Xiong [31] where they reported the available experimental and theoretical results on the thermodynamic properties of this system. However, in case of Ni-Ga binary system detailed thermodynamic assessment was reported by Yuan et al.[32] and Cao et al. [33] . But the phase diagrams proposed by Yuan et al. [32] and Cao et al. [33] have slight inconsistencies, which are discussed in subsequent sections. For Ni-Ga binary system, sufficient experimental results were available to define all the composition ranges and phase transitions of different phases. Moreover, the homogeneity ranges of some interesting and important phases like $Ni_3Ga$ and NiGa have been reassessed in this work. Mn-Ga and Ni-Ga binary systems are consistently used to design Ni-Mn-Ga based Heusler alloys for MCM applications. The target phase in most of these studies was reported to be FCC(γ). However due to lack of a consistent thermodynamic database for Ni-Mn-Ga system, the optimized compositions and the homogenization temperature for obtaining a single FCC(γ) phase could not be determined. Moreover, there are various intermetallic compounds and solution phases ($Ni_3Ga$, γ-Ni, NiGa, MnGa, γ-Mn etc.) in the Ni-Ga and Mn-Ga binaries, which can be explored for MCM applications. With the aid of a self-consistent thermodynamic database, the role of Mn and Ga content in Heusler alloys can be estimated and used for better MCM alloy design. Mn content in Heusler alloys controls the alloy's magnetism and, consequently, the second order magnetic transition of the alloys, whereas Ga content lowers the first order phase transition temperature [34]. Thus, by optimizing both Mn and Ga composition, new Heusler alloys can be developed based on the required properties.

CALPHAD (CALculation of PHAse Diagram) approach has been used in the present thermodynamic reassessment of the Ni-Ga, and Mn-Ga binaries. This is the first study to comprehensively evaluate the Mn-Ga binary system. Gibbs energy of liquid phase in both the binaries was described using Modified Quasichemical Model (MQM) [35,36]. In addition, the Compound Energy Formalism (CEF) was used to define Gibbs energy of solid solution [37]. In order to resolve the inconsistencies in the available phase diagram data for this system, few alloy compositions were selected to determine the phase transitions. By using the current optimized database for Mn-Ga and Ni-Ga system, an isothermal section for Ni-Mn-Ga is reported along with the potential composition region for MCM applications. This study is part of a comprehensive work to develop a thermodynamic database for Mn-Ni-Ga-Cu system for magnetocaloric applications.

## 2. Thermodynamic models and their parameters

All the calculations and optimization in the present study were performed using FACTSAGE thermochemical software [38,39]. The Gibbs free energy of all phases of pure Mn, Ni and Ga were taken from Dinsdale [40].

### 2.1 Pure elements and stoichiometric compounds

The Gibbs free energies of all the pure elements were taken from Scientific Group Thermodata Europe (SGTE) database [40] and the Gibbs free energies of pure compounds were determined based on the available experimental data such as heat capacity, standard enthalpy and entropy at 298.15 K. As per thermodynamic principle, the Gibbs free energies of pure elements can be calculated as follows:

$$G_T^{\circ} = H_T^{\circ} - TS_T^{\circ} \quad (4)$$

$$H_T^{\circ} = H_{298\,K}^{\circ} + \int_{298}^{T} C_p dT \quad (5)$$

$$S_T^{\circ} = S_{298\,K}^{\circ} + \int_{298}^{T} \left(\frac{C_p}{T}\right) dT \quad (6)$$

where $H_{298\,K}^{\circ}$ and $S_{298\,K}^{\circ}$ are the standard enthalpy and entropy at 298 K respectively and $C_p$ is the heat capacity at constant pressure. Data of fitting heat capacity was used to find heat capacity expression. In case of unavailable or not reliable data on $C_p$ values, the Neumann-Kopp rule [41] was used to predict $C_p$ values.

### 2.2 Liquid Solution

Modified Quasichemical Model (MQM) [35,36] was used to describe the Gibbs energy of liquid solution. MQM explains the short-range ordering (SRO) in the liquid phase and provides a more realistic description as compared to the conventional Bragg–Williams Random Mixing Model for entropy of the solution [36]. Instead of using a component fraction, a polynomial of a pair fraction can be used in the MQM to express the Gibbs energy of a pair formation. Additionally, the coordination number can change with composition. These MQM features give you more options for reproducing binary experimental data and combining optimized binary liquid parameters into a larger database for a multicomponent system. The reaction of pair exchanging in a binary A-B liquid solution can be expressed in the MQM by the distribution of A and B atoms over the quasi-lattice sites as follows.

$$(A-A)+(B-B)=2(A-B);\ \Delta g_{AB} \tag{7}$$

where (*i*-*j*) = First-Nearest Neighbour (FNN) pair between *i* and *j* components and $\Delta g_{AB}$ = Gibbs energy change of forming 2 moles of (*A*-*B*) pairs[36]. The following equation can be used to calculate the liquid solution's Gibbs energy:

$$G_L^{sol} = n_A g_A^{\circ} + n_B g_B^{\circ} - T\Delta S^{config} + \left(\frac{n_{AB}}{2}\right)\Delta G_{AB} \tag{8}$$

where the moles of *A*, *B*, and (*A*-*B*) pair are represented as $n_A$, $n_B$ and $n_{AB}$ respectively. $g_A^{\circ}$ molar Gibbs energies of the pure A and $g_B^{\circ}$ molar Gibbs energies of the pure B component. $\Delta S^{config}$ is the configurational entropy of random mixing (A-A), (B-B) and (A-B) pairs.

$$\Delta S^{config} = -R(n_A \ln X_A + n_B \ln X_B) - R\left(n_{AA}\ln\left(\frac{X_{AA}}{Y_A^2}\right) + n_{BB}\ln\left(\frac{X_{BB}}{Y_B^2}\right) + n_{AB}\ln\left(\frac{X_{AB}}{2Y_AY_B}\right)\right) \tag{9}$$

where $n_{ij}$ is the number of moles of (*i*-*j*) pairs. Pair fraction $X_{ii}$, mole fraction $X_i$, and the coordination equivalent fraction $Y_i$ are defined as follows:

$$X_{ii} = \frac{n_{ii}}{(n_{AA}+n_{BB}+n_{AB})} \tag{10}$$

$$X_A = \frac{n_A}{(n_A+n_B)} \tag{11}$$

$$Y_A = \frac{Z_A n_A}{(Z_A n_A + Z_B n_B)} \tag{12}$$

$$Z_i n_i = 2n_{ii} + n_{AB} \tag{13}$$

Where $Z_A$ and $Z_B$ are the coordination number of A and B respectively. Above equation was obtained for mass balance ((*i*-*j*) pairs). The model parameter $\Delta g_{AB}$ is expanded as a polynomial mentioned below.

$$\Delta g_{AB} = \Delta g_{AB}^{\circ} + \sum_{i,j\geq 1}\left(\Delta g_{AB}^{i0} X_{AA}^{i} + \Delta g_{AB}^{0j} X_{BB}^{j}\right) \tag{14}$$

Model parameters as mentioned in equation 15, $\Delta g_{AB}^{o}$, $\Delta g_{AB}^{io}$ and $\Delta g_{AB}^{jo}$ are function of the temperature:

The composition dependent coordination numbers are expressed as following:

$$\frac{1}{Z_A} = \frac{1}{Z_{AA}^{A}}\left(\frac{2n_{AA}}{2n_{AA}+n_{AB}}\right) + \frac{1}{Z_{AB}^{A}}\left(\frac{2n_{AB}}{2n_{AA}+n_{AB}}\right) \quad (15)$$

$$\frac{1}{Z_B} = \frac{1}{Z_{BB}^{B}}\left(\frac{2n_{BB}}{2n_{BB}+n_{AB}}\right) + \frac{1}{Z_{BA}^{B}}\left(\frac{2n_{AB}}{2n_{AA}+n_{AB}}\right) \quad (16)$$

where $Z_{AA}^{A}$ and $Z_{AB}^{A}$ represents $Z_A$ value when all the nearest-neighbour atoms of atom A are A and Bs respectively. $Z_{BB}^{B}$ and $Z_{BA}^{B}$ are defined analogously. The ratio of the coordination numbers ($Z_A/Z_B$) estimates the composition of the maximum SRO in each binary subsystem. In the present study, $Z_A/Z_B$ was set to 6 for liquid phase of both the binaries.

## 2.3. Solid solutions

In the Mn-Ga and Ni-Ga system, Gibbs energies of the solid solutions (such as **γ** and $Mn_3Ga$ phases) were determined using Compound Energy Formalism (CEF)[42] . Solid solution comprises of two kinds of sublattices such as interstitial and substitutional sites $(A, B)_n\,(C, D)_m$ in CEF. The Gibbs free energy of a solid solution can be written as:

$$G_{Sol}^{S} = Y_A' Y_A'' G_{A:C} + Y_A' Y_D'' G_{A:D} + Y_B' Y_C'' G_{B:C} + Y_B' Y_D'' G_{B:D} + nRT(Y_A' lnY_A' + Y_B' lnY_B') + mRT(Y_C'' lnY_C'' + Y_D'' lnY_D'') + \sum_{i,j,k} Y_i' Y_j' Y_K'' L_{ij:k} + \sum_{i,j,k} Y_k' Y_i'' Y_j'' L_{k:ij} \quad (17)$$

where $Y_i'$ and $Y_i''$ are the site fraction of species *i* in the corresponding sublattice. $G_{i:j}$ is the Gibbs free energy of an end-member, the $L_{i,j:k}$ and $L_{k:i,j}$ are the interaction parameters between the component *i* and *j* on one sub lattice when the other sub lattice is occupied by k.

While determining the sublattice models for solution phases in the Mn-Ga system various discrepancies in the crystal structure have been observed. $Mn_3Ga$ phase is reported with three types of crystal structures, namely tetragonal [43,44], cubic [45,46] and hexagonal [47–49]. According to Masumoto et al. [43] and Niida et al. [44] tetragonal $Mn_3Ga$ with $D0_{22}$ ordering precipitated from γ-Mn phase after annealing the alloys at 350-400 °C. Cubic structure of $Mn_3Ga$ was predicted using DFT (Density Functional Theory) calculations [45,46] which was later experimentally confirmed by Kharel et al.[50]. Kharel et al. [50] used high temperature XRD and thermal analysis to report cubic $Mn_3Ga$ transformation to tetragonal $Mn_3Ga$ at 327 °C and further transformation to hexagonal $Mn_3Ga$ at 527 °C. In addition, hexagonal $Mn_3Ga$ with $D0_{19}$ ordering was also observed in water quenched alloys between 600 and 800 °C using XRD [47–49]. In present study, hexagonal $Mn_3Ga$ ($P6_3/mmc$) was considered while optimizing and sublattice model for $Mn_3Ga$ is based on their two Wyckoff sites mentioned in table 1. Similarly, various studies reported different stable crystal structure for MnGa [47,51–53]. Schubert et al. [47] reported the hexagonal structure of MnGa with space group R3m. Minakuchi et al. [51]

determined the crystal structure of MnGa as monoclinic using XRD. In addition, an AuCu type structure with $L1_0$ ordering was predicted for MnGa through DFT calculations by Sakuma and Yang et al. [52,53]. In the present study we have adopted two sublattice formulism corresponding to two Wyckoff sites as mentioned in Table 1. Tsuboya and Sugihara [54] reported the crystal structure of $Mn_8Ga_5$ as γ- brass type structure. However, Meissner et al. [55] determined that the crystal structure is not perfectly γ- brass type structure due to presence of reflection pattern from a hexagonal lattice. Minakuchi et al. [51] reported crystal structure of low temperature $Mn_8Ga_5$ phase similar to $Al_4Cu_9$ which was also observed by Tozman et al. [56]. In their study, Tozman et al. [56] determined crystal structure of $Mn_8Ga_5$ similar to a layered tetragonal unit cell with $L1_0$ ordering with the help of DFT calculations and experiments. They reported eight possibilities of atomic arrangement according to Wyckoff sites. In addition, they provided (Mn, Ga), (Mn, Ga), (Mn), (Mn, Ga) as the sublattices to explain these Wyckoff positions in $Mn_8Ga_5$. The same sublattice formulism suggested by Tozman et al. [56] is adapted in present thermodynamic assessment. The crystal structure of $Mn_2Ga$ (r) (Room Temperature) is AuCu type with the space group *P4/mmm* reported by Meissner et al. [55]. The two sublattice model formulism $(Mn, Ga)_2$ (Mn, Ga) is used for $Mn_2Ga$ (r) in the present study based on their two Wyckoff sites. Meissner et al. [55] determined the crystal structure of the $Mn_2Ga$(h) (High Temperature) phase to be Mg type with *P63/mmc* space group. The two sublattice model formulism $(Mn, Ga)_2$ (Mn, Ga) is used in the present thermodynamic assessment corresponding to their two Wyckoff sites. The crystal structure $Mn_3Ga_2$ was reported to be AuCu type with *P4/mmm* space group by Minakuchi et al. [51]. In the present work, $Mn_3Ga_2$ is treated as the formula $(Mn, Ga)_3$ $(Mn, Ga)_2$ by a two sublattice model based on their two Wyckoff sites. Apart from crystallographic information, the magnetic properties of a few intermetallic compounds of Mn-Ga binary system are included in the present study as well. Tsuboya and Sugihara [57,58] studied the magnetic behaviour of various intermetallic compounds of Mn-Ga binary system and determined the magnetic transition temperature as well as mean magnetic moment ($\mu_B$/Mn) of $Mn_3Ga$ [54], $Mn_8Ga_5$ [54] and $Mn_3Ga_2$ [58]. Wachtel and Nier [59] reported the magnetic transition temperature of $Mn_2Ga$(r) while Bither and Cloud [60] determined the magnetic moment of the same phase. All the magnetic properties discussed above were considered while optimising the different magnetic phases in the present study (tabulated in Table 2). On the other hand, the solid solutions in the Ni-Ga system are well defined by the Yuan et al. [32] and Cao et al. [33]. In the present study same sublattices for each solid solution has been adopted. In addition, Mn-Ni system was also reassessed to make the database consistent. The liquid phase of Mn-Ni was described using the MQM. Optimised parameters for the liquid phase of the Mn-Ni system are also provided in Table 2, while all other parameters and sublattices were kept as described by Guo and Du [61].

Using the above-mentioned solution models, the thermodynamically optimized parameters for different phases in Mn-Ga and Ni-Ga binary systems are summarised in Table 2. Similarly, the thermodynamically optimized parameters for different intermetallic compounds of both binaries are tabulated in Table 3.

## 3. Literature Survey

### 3.1 Mn-Ga binary system

Crystallographic, magnetic and structural data of the various solid phases is summarised and explained in detail by Hao and Xiong [31]. First experimental study of Mn-Ga phase diagram was performed by Meissner et al. [55]. They reported thirteen invariant reactions and ten intermetallic compounds for the whole composition range of Mn-Ga phase diagram using XRD and thermal analysis. Later, Wachtel and Nier [59] determined the phase diagram based on magnetic susceptibility measurements. They reported thirteen intermetallic compounds using electronic and magnetic properties of Mn-Ga solid phases. Using X-ray diffraction investigation, Lu et al. [63] reported Mn-Ga phase diagram with five intermetallic compounds as $Mn_2Ga_5$, $MnGa_4$, $MnGa_5$ and $Mn_3Ga$. Recently, Mn-rich side of the Mn-Ga phase diagram was investigated by Minakuchi et al. [51] and reported $Mn_2Ga$(H), $Mn_8Ga_5$, $Mn_7Ga_6$, $Mn_2Ga$(R) and $Mn_3Ga_2$ phases in the phase diagram. They reported phase equilibria upto the 50 at.% Ga using the Electron Probe Micro Analysis (EPMA) and thermal analysis of the alloys. They reported same phases to that of Wachtel and Nier [59] except the difference in temperature of formation and homogeneity range of $Mn_3Ga$ and $Mn_7Ga_6$ phases. Meissner et al. [55] reported the δ-Mn phase with a solubility of less than 10 at.% Ga. They also reported the formation of γ-Mn by the peritectic reaction between δ-Mn and liquid at 1453 K, and formation of $Mn_2Ga$(H) from a peritectic reaction between γ-Mn and liquid at 990 K. The phase diagram reported by Wachtel and Nier [59] described the homogeneity range of δ-Mn till 30 at.% Ga with δ-Mn metatectically decomposing into γ-Mn and liquid at 1287 K. Experimental results reported by Meissner et al. [55] and Wachtel and Nier [59] were found consistent with each other for the peritectic reaction at 990 K. In addition, Lu et al. [63] reported a wider homogeneity for δ-Mn (0-0.45 at.% Ga within a temperature range of 1517.15 to 988.15 K) and γ-Mn phase(0-0.22 at.% Ga within a temperature range of 1406.15 K to 893.15 K). They also reported polymorphic transformations in γ-Mn phase. They reported the high temperature face centred cubic phase ($\gamma_1$) transformation to face centred tetragonal phase ($\gamma_2$) at around 1093.15 K, which transformed to a face centred tetragonal phase ($\gamma_3$) with long range order at 988.15 K. They reported $\gamma_3$ phase is stable at room temperature with homogeneity range of 0.37 to 0.45 at.% Ga which was not reported by any other study[51,55,59]. However, on the Ga rich side, they reported only three intermetallic compounds ($Mn_2Ga_5$, $MnGa_4$ and $MnGa_5$). Findings of Minakuchi et al.[51] about the δ-Mn also showed good agreement with the findings of Wachtel and Nier [59] but former reported a higher temperature (1313 K) for the peritectic reation between $Mn_2Ga$(H), δ-Mn and liquid. $Mn_3Ga$ compound was not reported in the study by Meissner et al. [55]. On the other hand, Wachtel and Nier [59] reported a peritectoid reaction between $Mn_3Ga$, β-Mn and $Mn_2Ga$(R) at 783 K with a narrow homogeneity range. Lu et al. [63] also suggested that $Mn_3Ga$ forms due to congruent transformation from γ-Mn at 1093 K. Minakuchi et al.[51], reported $Mn_3Ga$ formation by a peritectoid reaction between γ-Mn and $Mn_2Ga$(H) at 1003 K. Other discrepancies include the homogeneity ranges of different Mn-Ga phases and the differences in solubilities of intermetallic compounds. According to Meissner et al. [55], MnGa forms by a peritectic reaction between liquid and $Mn_8Ga_5$ while Wachtel and Nier [59] reported that reaction involves $Mn_7Ga_6$ instead of $Mn_8Ga_5$ compound.

In addition, the enthalpies of formation for the intermetallic compounds in the system was calculated by different authors using density function theory (DFT) at 0 K [64–67]. Sedmidubský et al.[68] used CALPHAD approach to estimate the enthalpies of formation of all intermetallic compounds at 298 K. Miedema et al.[69] used the Miedema model to calculate the enthalpies of formation for intermetallic compounds at 298 K. In addition, Winterlik et al. [70] measured the Cv values for $Mn_2Ga$ and $Mn_3Ga$ alloys while heating the samples from 0 to 300 K. To the best of authors' knowledge, no experimental data regarding thermodynamic properties is available for Mn-Ga system except $C_v$ values for $Mn_2Ga$ and $Mn_3Ga$. There is very limited work on the thermodynamic optimization for the Mn-Ga system. One of the preliminary attempt was carried by Sedmidubsky et al.[68], where they considered liquid phase as an ideal solution, and intermetallic compounds as stoichiometric compounds. They have not compared the calculated phase diagram with the available experimental data. To determine the thermodynamic behaviour of liquid in the system, they consider regular solution model: Redlich-Kister equation (RK). Moreover, they also used ab-initio electronic structure calculations of cohesive energies to evaluate the enthalpy of formation of intermetallic compounds. To the best of authors' knowledge, no other work has been carried out for the thermodynamic optimization of the Mn-Ga system.

### 3.2 Ni-Ga binary system

First phase diagram study of the system was carried out by Hellner et al. [71]. Using X-ray diffraction, thermal analysis and microscopy, they found γ-Ni, NiGa and, $Ni_3Ga_2$ solid solutions, and $Ni_{13}Ga_9$, $Ni_5Ga_3$, $Ni_2Ga_3$, and $NiGa_4$ stoichiometric compounds in the system. Solubility of Ga in γ-Ni was reported to be 28 at.% at 1484 K. $Ni_3Ga$ ($L1_2$) ordered superstructure of γ-Ni was first reported by Pearson [72]. Later, Pearson and Rimek [73] reassessed the system and found the maximum solubility of Ga in γ-Ni phases is 24.3 at.% at 1485 K. XRD, DTA, and EPMA analysis of the various compositions was carried out by Feschotte et al. [74]. They reported $Ni_3Ga_2$ to $Ni_{13}Ga_9$ transformation at 1063 K. NiGa phase boundaries were determined by Isper et al.[75] in which they reported higher Ga solubility in the NiGa phase as compared to that of Feschotte et al.[74]. In addition, his work was mainly focused on Ga-rich side of the phase diagram. In order to determine the phase equilibria in Ni-rich side of the phase diagram, Micke et al.[76] analysed the alloys using DTA and diffusion couple analysis. They reported four invariant reactions at 1222, 1004, 948, and 1048 K in the Ni-rich side of the system. In addition, Ikeda et al. [28] determined the high temperature solubility of the $Ni_3Ga$ phase within the temperature range of 1073 to 1373 K. They reported a narrow homogeneity range of Ga in the $Ni_3Ga$ as compared to that of Feschotte et al.[74]. Recently Schmetterer et al.[29] and Ducher et al.[30] reported studies on Ga-rich and Ni-rich sides of the phase diagram respectively. In Ni-rich side, Ducher et al. [30] reported similar invariant reaction as reported by Micke et al. [76] but at slightly different temperatures (1215, 1000, 992 and 1065 K). They also reported congruent melting of $Ni_3Ga_2$ and $Ni_{13}Ga_9$ at 35 at.% and 41 at.% Ni respectively. Meanwhile, Schemtterer et al. [29] found $Ni_3Ga_7$ to be the most stable phase in the Ga rich region of the phase diagram. In their thermodynamic optimization, Yuan et al. [32] considered $NiGa_4$ as most stable compound on the contrary, Cao et al. [33] considered $Ni_3Ga_7$ as the most stable compound in the Ga rich side of the system. In addition, in the Ni-rich side of the phase diagram Yuan et al. [32] considered $Ni_5Ga_3$ and $Ni_3Ga_2$ as line compounds but in the later studies it was confirmed that there exists homogeneity in both the phases at high temperature. Meanwhile, Ducher et al. [30] also reported $Ni_3Ga_2$ is a high temperature phase. In addition, the compound $Ni_{13}Ga_9$ was reported in this system[76]. All the phase equilibria and the homogeneity ranges in Ni rich phases were well incorporated in the thermodynamic assessment by Cao et al. [33].

Enthalpies of formation at 300 K were measured using solution calorimetry by Martosudirjo and Pratt [77]. Jacobi et al. [78] measured enthalpy of formation at 1023 K using solution calorimetry as well. Predel et al. [79] also used same technique to measure the enthalpy of formation and EMF technique to measure the Gibbs free energy of formation. Pratt and Bird [80] measured the thermodynamic properties (activity, integral Gibbs energy, enthalpies and entropies) of $\gamma$-Ni solid solution and the intermetallic phases of the system within a temperature range of 873-1100 K using EMF method. Activity of Ga at 1073-1273 K in 5 at.% to 58 at.% Ga alloys was measured using EMF technique by Katayama et al. [81]. Seybolt [82] and Mikula et al. [83] measured activity of Ga in NiGa at 1108 K and 1173 K, and 1173 K respectively. Seybolt [82] also measured the partial Gibbs energy of Ga in NiGa at 1108 and 1173 K. Meschel and Kleppa [84] used high temperature direct synthesis calorimetry to measure standard enthalpy of formation of NiGa at 1373 K. Activity of Ga in $Ni_3Ga$ was measured by Yuan et al. [85] and Kushida et al. [86] using EMF technique. Mixing enthalpy of liquid phase at 1661, 1594, 1534, 1519, 1407 and 1304 K was measured using drop calorimetric technique by Schmetterer et al.[29].

Ni-Ga was thermodynamically assessed first by Yuan et al. [10] and then by Cao et al. [9]. Both used regular solution model to explain the liquid solution but the reported phase diagrams had various differences in phase equilibria and homogeneity ranges. Yuan et al. [9] considered a lower liquidus for Ga-rich side, which was later experimentally determined by Feschotte et al. [11]. Cao et al. [9] incorporated these experimentally determined liquidus values. Yuan et al. [9] considered $NiGa_4$ as a Ga rich stoichiometric compound while Cao et al. [9] considered $Ni_3Ga_7$ as the most stable stoichiometric compound in the Ga- rich region of Ni-Ga system. Moreover, the recent studies by Ducher et al. [12] and Schmetterer et al. [13], carried out after the thermodynamic assessment of Yuan et al. [10], were considered by Cao et al. [9]. Yuan et al. [10] did not consider the existence of $Ni_3Ga_2$ phase while Cao et al reproduced it, as its existence was validated by various studies [11, 12, 14, 15].

## 4. Experimental details

To resolve the existing discrepancies in the Mn-rich region of Mn-Ga system, in the present work, compositions of Mn-rich region ($Mn_{0.95}Ga_{0.05}$, $Mn_{0.88}Ga_{0.12}$, $Mn_{0.82}Ga_{0.18}$ and $Mn_{0.75}Ga_{0.25}$) were investigated. The primary target was to determine the phase equilibria among δ, γ and $Mn_3Ga$ phases in the Mn-Ga system. In addition, to determine the liquidus temperature in the Ga-rich region of Ni-Ga phase diagram alloy with 0.9 at.% Ga was developed. In addition, NiGa phase was reinvestigated by developing $Ni_{0.5}Ga_{0.5}$ and $Ni_{0.47}Ga_{0.53}$ alloys in in present work. Mn, Ni and Ga with 99.9% purity were used to cast the alloys in tungsten inert gas arc melting unit. Due to the oxidising nature of Mn, melting chamber was first evacuated ($10^{-5}$ bar) and then filled back with argon gas before melting. In order to ensure the homogeneity, each alloy was melted at least 3 times. Composition of the as cast alloys was determined using PAN Analytics X-Ray fluorescence (XRF) instrument. Thermal analysis of the as cast alloys (weighing~50-100 mg) was carried out in NETZSCH STA Jupiter 449 F3 thermal analyser under Ar atmosphere at a heating rate of 10 K/min.

## 5. Results and discussion

### 5.1 Mn-Ga binary system

For the current thermodynamic assessment, experimental studies of Wachtel and Nier [59], and Minakuchi et al. [51] were given priority over experimental studies reported by Meissner et al.[55] and Lu et al.[63]. As shown in Fig. 1, both the studies[51,59] shows similarities such as existence of $Mn_7Ga_6$ (not reported by Meissner et al.[55] and Lu et al.[63]), homogeneity of δ-Mn (reported very narrow by Meissner et al. [55]), existence of $Mn_2Ga$(R) and $Mn_2Ga$(H) (not reported by Lu et al.[63]), etc. Seven intermetallic compounds ($MnGa_6$, $MnGa_4$, $Mn_2Ga_5$, $Mn_3Ga_5$, $Mn_5Ga_7$, $Mn_5Ga_6$ and $Mn_7Ga_6$) were considered in the present work while $Mn_3Ga$, $Mn_2Ga$(R), $Mn_2Ga$(H) and $Mn_3Ga_2$ were treated as solid solutions to explain the experimental data. The calculated phase diagram as shown in Fig. 1 is in good agreement with the experimental data provided by Minakuchi et al. [51] for 0 to 50 at.% Ga. As discussed, there are discrepancies among the experimental studies in the Mn-rich region. To authenticate the phase equilibria in the Mn rich region of the phase diagram, DTA of the as cast samples was carried out in the present study. As shown in Fig. 2(a), $Mn_{0.75}Ga_{0.25}$ showed the first thermal arrest as an endothermic peak at 974.7 K and a sharp liquidus peak at 1317.9 K, which is slightly lower as compared to that reported by Minakuchi et al. [51](shown in Fig. 3). As the Mn proportion was increased, the liquidus can been seen shifting to higher temperature values. As shown in Fig. 2(b) $Mn_{0.82}Ga_{0.12}$ reports endothermic peaks at 1015.4 K, 1361 K, and 1450.1 K.

Endothermic peak at 1015.4 K corresponds to the β+$Mn_3Ga$↔γ phase equilibria, while peak at 1361 K corresponds to γ-Mn to δ-Mn transformation. In addition, the endothermic peak at 1450.1 K reflects the liquidus temperature for the alloy. Similarly, In Fig. 2(c) a thermal arrest at a temperature of 1474.7 K denotes the liquidus for $Mn_{0.88}Ga_{0.12}$ alloy. At 1053.4 K and 1440.1 K there were changes in the DTA curve, which can be attributed to the dissolution of β phase to β+γ phase equilibria, and δ to δ+L phase equilibria respectively. For $Mn_{0.95}Ga_{0.05}$, first endothermic peak (shown in Fig. 2(d)) observed during heating was at 1201.8 K, corresponds to $\beta \leftrightarrow \gamma+\beta$ phase equilibria. The liquidus for this alloy is at 1531 K as shown in Fig. 2(d). The thermal analysis of the investigated as cast alloys showed reasonable agreement with already reported experimental data as well as the calculated phase diagram using the present thermodynamically optimized parameters (shown in Fig. 3). Particularly the experimental data obtained from DTA analysis in present study was cosistent with the experimental data reported by Minakuchi et al. [51] while, the liquidus in the Ga rich region agrees with experimental data reported by Wachtel and Nier [59] (shown in Fig. 3).

Minakuchi et al. [51] considered $Mn_7Ga_6$ as a solid solution with homogeneity (0.45-0.50 at.% Ga) at 994.15 K while Wachtel and Nier [59] reported very narrow homogeneity range (0.45-0.46 at.% Ga) at 1045.15 K. On the contrary, Meissner et al. [55] and Lu et al. [63] did not report the existence of this compound. In present thermodynamic optimization, $Mn_7Ga_6$ was treated as intermetallic compound for simplicity. Minakuchi et al. [51] have also reported a higher homogeneity 0.32-0.36 at.% Ga for $Mn_2Ga$(R) phase at 773.15 K, which was not considered in the present study. Also the homogeneity range reported for $Mn_2Ga$(H) phase was 0.3-0.35 at.% Ga at 1183.15 K [51] which differs to other reported studies[55,59]. While Lu et al. [63] did not report the existence of both $Mn_2Ga$(R) and $Mn_2Ga$(H) phases in the Mn-Ga system. In the present optimization, existence of both the phases was considered and the homogeneity ranges of both the low and high temperature phases of $Mn_2Ga$ phase were explaining the experimental data of Wachtel and Nier [59], and Minakuchi et al. [51]. As shown in Fig. 3, Wachtel and Nier [59] did not provide any information about the homogeneity ranges but provided the liquidus information of the Mn-Ga system. In the present thermodynamically assessed phase diagram, β phase differ from the experimental studies in terms of homogeneity range at low temperature as well as the two-phase region (β + γ). Apart from this, Ga- rich side was thermodynamically optimized to explain the experimental data of Wachtel and Nier [59] (shown in Fig. 3). So in present thermodynamic assessment authors' considered seven intermetallic compounds ($MnGa_6$, $MnGa_4$, $Mn_2Ga_5$, $Mn_3Ga_5$, $Mn_5Ga_7$, $Mn_5Ga_6$ and $Mn_7Ga_6$) while $Mn_3Ga$, $Mn_2Ga$(R), $Mn_2Ga$(H) and $Mn_3Ga_2$ as solid solutions to explain the reported experimental data[51,55,59,63]. Figure 4(a) shows the calculated phase equilibria between γ, liquid (L), δ and $Mn_2Ga$(H) phases along with the reported experimental data[51,59]. Figure 4(b) and (c) shows the high temperature homogeneity range of $Mn_3Ga$ and $Mn_2Ga$(R) phases, and $Mn_3Ga_2$ and MnGa phases respectively.

Calculated thermodynamic properties for Mn-Ga system is summarized in Fig. 5. Enthalpy of formation of intermetallic compounds at 298 K was calculated with Mn(s) and Ga (s) as reference (shown in Fig. 5(a)). The calculated enthalpies shows a reasonable agreement with that of estimated values by Miedema et al. [69] at 298 K. There is no other experimentally determined thermodynamic properties available for this system. Using the thermodynamically optimized database, activity of the Mn and Ga in the liquid phase was calculated at 1573 K as seen in Fig. 5 (b). The calculated mixing enthalpy of liquid at 1573 K shows a minimum at approximately 0.42 at.% Ga (shown n Fig. 5(c)) suggesting short range ordering in liquid phase at that composition and temperature.

**5.2 Ni-Ga binary system**

The thermodynamically reassessed phase diagram of Ni-Ga along with experimental data is shown in Fig. 6. Solid lines represent the calculated phase regions while symbols represent the experimental data. As compared to previous assessments of the system by Yuan et al.[32] and Cao et al.[33], there are some improvements in the thermodynamic modeling of this system which is mentioned below. As seen in Fig. 6, the liquidus shows better agreement with the experimental data in the entire composition range (shown in Fig. 6). The experiment data reported by Schmetterer et al.[29] in the Ga-rich side of the system was reproduced in the present thermodynamic optimization. Moreover, the invariant reaction in the Ga rich side (Liquid

+$Ni_2Ga_3$↔$Ni_3Ga_7$) was also reproduced and showed a good agreement with the thermodynamic assessment of Cao et al. [33]. In addition, the other invariant reactions as reported by Cao et al. [33] were also reproduced in the present study. Phase boundary between NiGa and the L+ NiGa region on the Ga rich side of the system shows better agreement with experiment data. In addition, homogeneity ranges of $Ni_3Ga$ and NiGa phases were found in better agreement with the experimental data as compared to earlier thermodynamic optimization[32,33]. The calculated enthalpy of formation at 298 K was found to be consistent with Cao et al. [33] (shown in Fig. 7(a)). Experimentally measured activity of Ga in Ni-Ga system by Pratt et al. [80] was reproduced reasonably well in the present study. Figure 7(b) shows the change in log ($a_{Ni}$) and log ($a_{Ga}$) with the Ga composition (in mole fraction) at 1100 K. Cao et al. [33] stated that, reproducing the liquidus in the phase diagram comprises the mixing enthalpy of the liquid phase, so they gave importance to mixing enthalpy data leading to lower liquidus in the phase diagram. But using the MQM model, authors were able to explain the mixing enthalpy of the liquid and liquidus temperatures in the present study. As seen in Fig. 8, the mixing enthalpy of the liquid at 1519 K and the liquidus temperature (shown in Fig. 6) are simultaneously reproduced and agree well with the experimental data.

Thermal analysis of the as cast alloys using DTA of the Ni-Ga alloys is shown in Fig. 9. For $Ni_{0.5}Ga_{0.5}$ alloy, the DTA curve shows an endothermic peak at 1474 K as shown in Fig. 9(a). This peak corresponds to the liquidus temperature for the $Ni_{0.5}Ga_{0.5}$ alloy. As Ga content was increased from 0.5 to 0.53 at.%, liquidus temperature was found shifting to a lower temperature value. As evident in DTA curve of as cast $Ni_{0.47}Ga_{0.53}$ alloy the liquidus temperature was 1444 K as denoted by endothermic peak in Fig. 9(b). In addition, alloy with 0.9 at.% Ga shows first endothermic peak at 671 K while heating. This endothermic peak corresponds to the dissolution of $NiGa_7$ and Ga(l) to $Ni_2Ga_3$ and liquid solution. The small change noticeable at 1034 K in the DTA curve can be possible due to complete dissolution of $Ni_2Ga_3$ to liquid phase. The onset temperature of all the as-cast alloys is incorporated in the Fig. 6 and compared with the calculated and experimental data. All the phase transitions found in the as cast alloys are supporting the phase transitions in the Ga-rich region of calculated phase diagram (shown in Fig. 6).

Present thermodynamic assessment of Mn-Ga and Ni-Ga will aid in developing various ternary systems such as Ni-Mn-Ga and Ni-Ga-As for MCM applications. Using the thermodynamically optimized parameters of the Mn-Ga and Ni-Ga binaries, along with previously optimized Mn-Ni system [61], a self-consistent thermodynamic database for Ni-Mn-Ga system is developed. A preliminary isothermal section at 1073 K is shown in Fig. 10. As discussed previously Mn and Ga compositions play an important role in determining the martensite to austenite transformation and magnetic phase change which ultimately determined the refrigeration capacity of the alloy. All the previous magnetocaloric compositions have targeted FCC (γ) (austenite) as the starting phase in order to obtain martensite phase upon quenching. The present database can aid in deciding the composition range of FCC (γ). As seen in Fig. 10, the composition range for the FCC (γ) phase at 1073 K is highlighted which can aid in selecting various alloys with varying Mn and Ga compositions.

## 6. Conclusion

Mn-Ga and Ni-Ga binaries were thermodynamically assessed in the present study using CALPHAD approach. The liquid solution of Mn-Ga and Ni-Ga binaries was described by Modified Quasichemical Model (MQM), which explains the short-range ordering (SRO) in the liquid solution. Solid solutions in both the binaries were described using the Compound Energy Formalism (CEF). As discussed in Mn-Ga system, phase regions of δ-Mn, γ-Mn, existence of $Mn_2Ga$(R), $Mn_2Ga$(H) and phase transition of $Mn_3Ga$ have discrepancies in the Mn-rich region. Using the thermal analysis of as cast alloys in present study, $Mn_3Ga$ to γ-Mn and $Mn_2Ga$(R) transformation was obtained at 974.70 K and liquidus at 1317.9 K for $Mn_{0.75}Ga_{0.25}$ composition. For $Mn_{0.82}Ga_{0.18}$, DTA result showed the β+$Mn_3Ga$↔γ at 1015.4 K phase equilibria, γ-Mn to δ-Mn transformation at 1361 K, and liquidus at 1450.10 K. For $Mn_{0.88}Ga_{0.12}$, 1053.4 K and 1440.1 denoted the β↔ β+γ, and δ ↔ δ+L phase equilibria. At 0.95 at.% Mn, β-Mn to γ-Mn transition at 1201.80 K while at 1531 K, δ-Mn to liquid phase transformation was obtained. Experimental results of present study ascertain the existence of $Mn_3Ga$ phase, wider homogeneity range of δ-Mn compared to Meissner et al. [55], phase equilibria of δ-Mn and similar liquidus temperature values to that of Minakuchi et al. [51] and Wachtel and Nier [59] at Mn rich side of the phase diagram. In addition, phase transitions determined using the DTA were also in good agreement with the calculated phase diagram of Mn-Ga system. In Ni-Ga system, the liquidus in the Ga-rich side was consistent with the experimental data as compared to earlier thermodynamic assessments[32,33]. In addition, the solubility of Ga in Ga-rich side of NiGa phase and liquidus of Ni-Ga system was re-defined to explain the experimental data. Moreover, the mixing enthalpy of the liquid phase at 1573 K for Ni-Ga system was found in good agreement with the experimental data as compared to earlier studies [32,89]. Using the current developed self-consistent thermodynamically optimized database of Mn-Ga and Ni-Ga, thermodynamic assessment of Ni-Mn-Ga ternary system (Heusler alloys) can be performed.

Figures

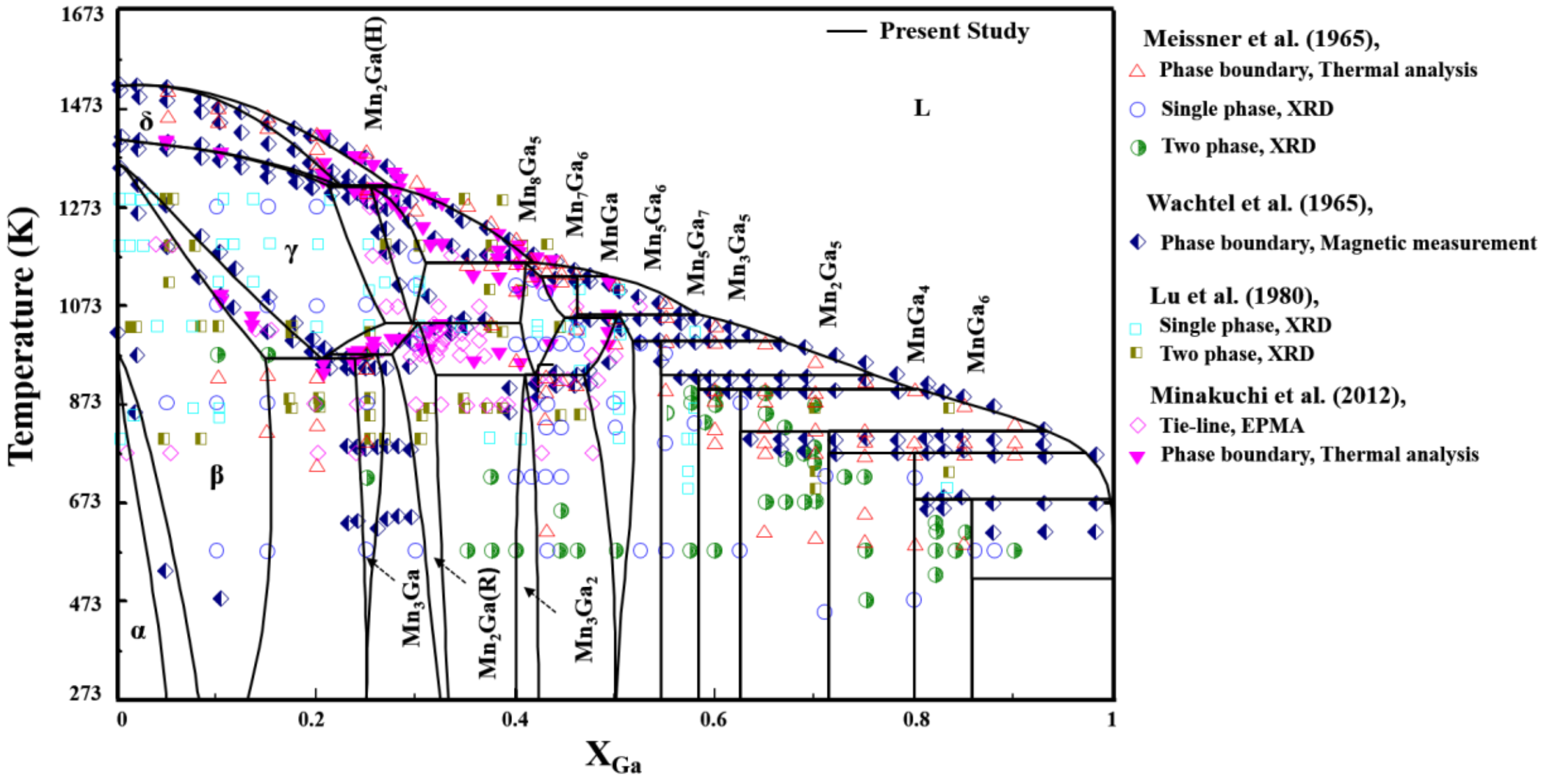

Fig. 1 Calculated phase diagram of Mn-Ga along with experimental data reported by Meissner et al. [55], Lu et al. [63], Wachtel and Nier [59], and Minakuchi et al. [51].

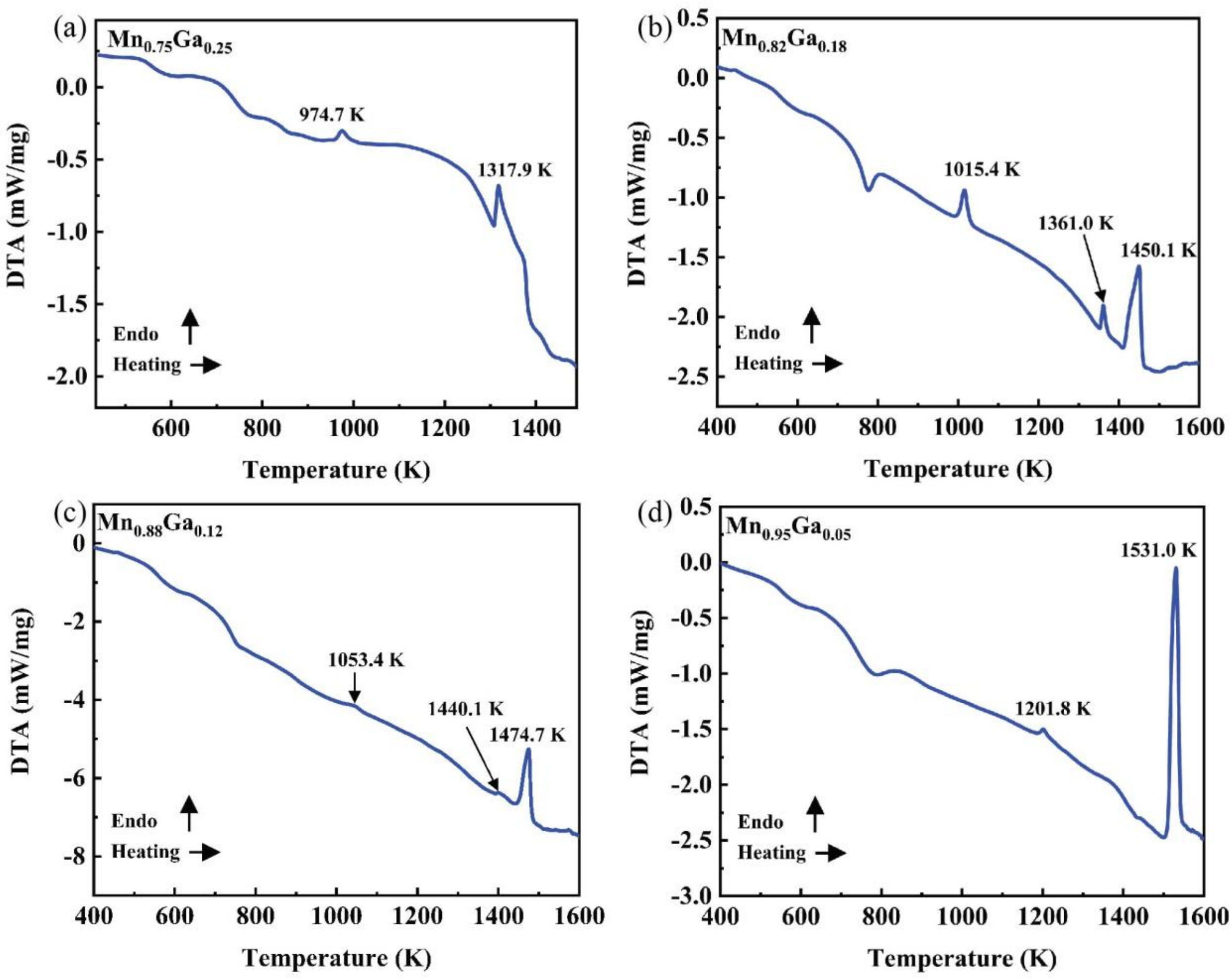

Fig. 2 DTA curves for (a) $Mn_{0.75}Ga_{0.25}$, (b) $Mn_{0.82}Ga_{0.18}$ (c) $Mn_{0.88}Ga_{0.12}$, and (d) $Mn_{0.95}Ga_{0.05}$ alloys.

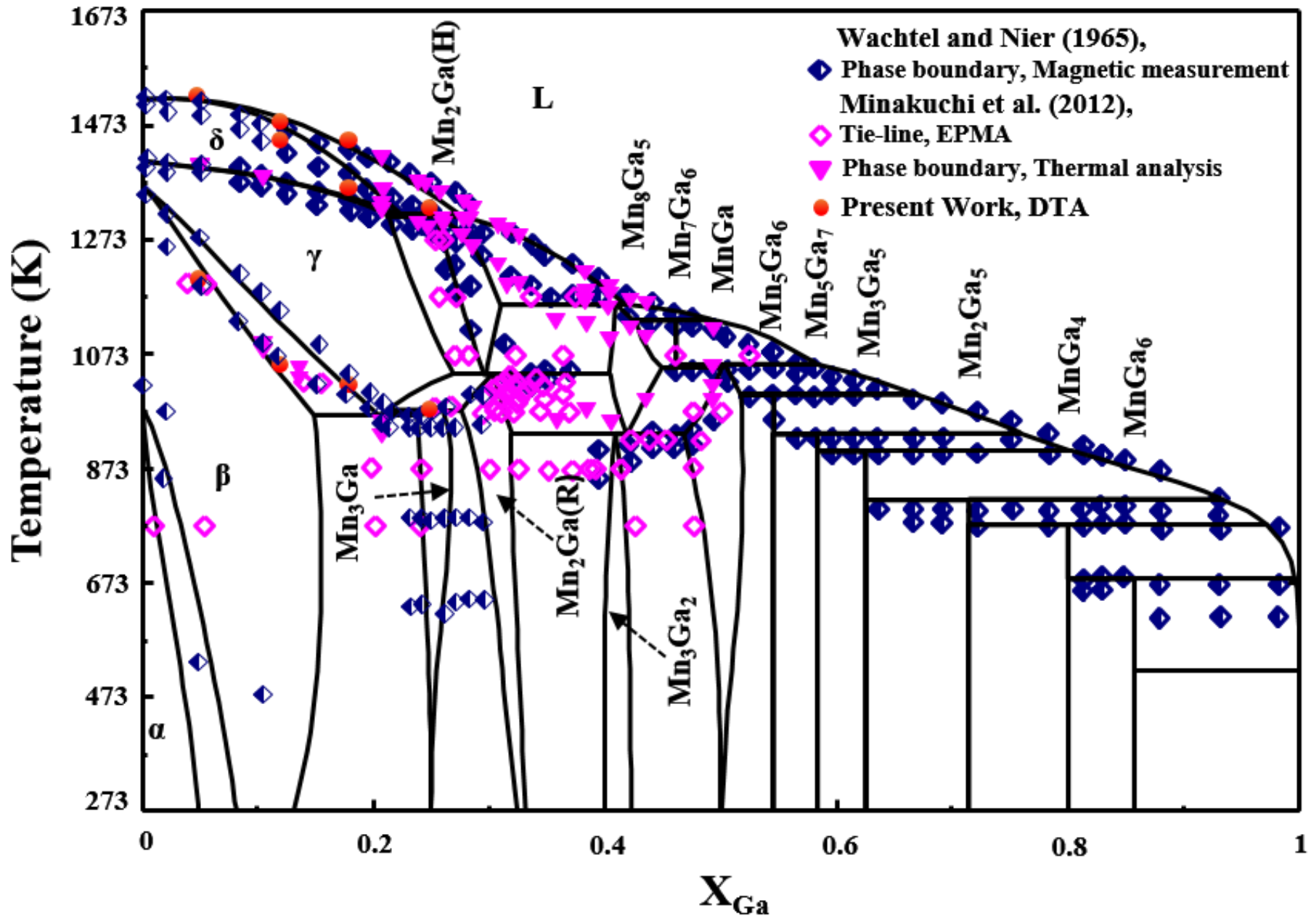

Fig. 3 Calculated phase diagram of Mn-Ga along with DTA analysis in comparison with experimental data provided by Wachtel and Nier [59] and Minakuchi et al. [51].

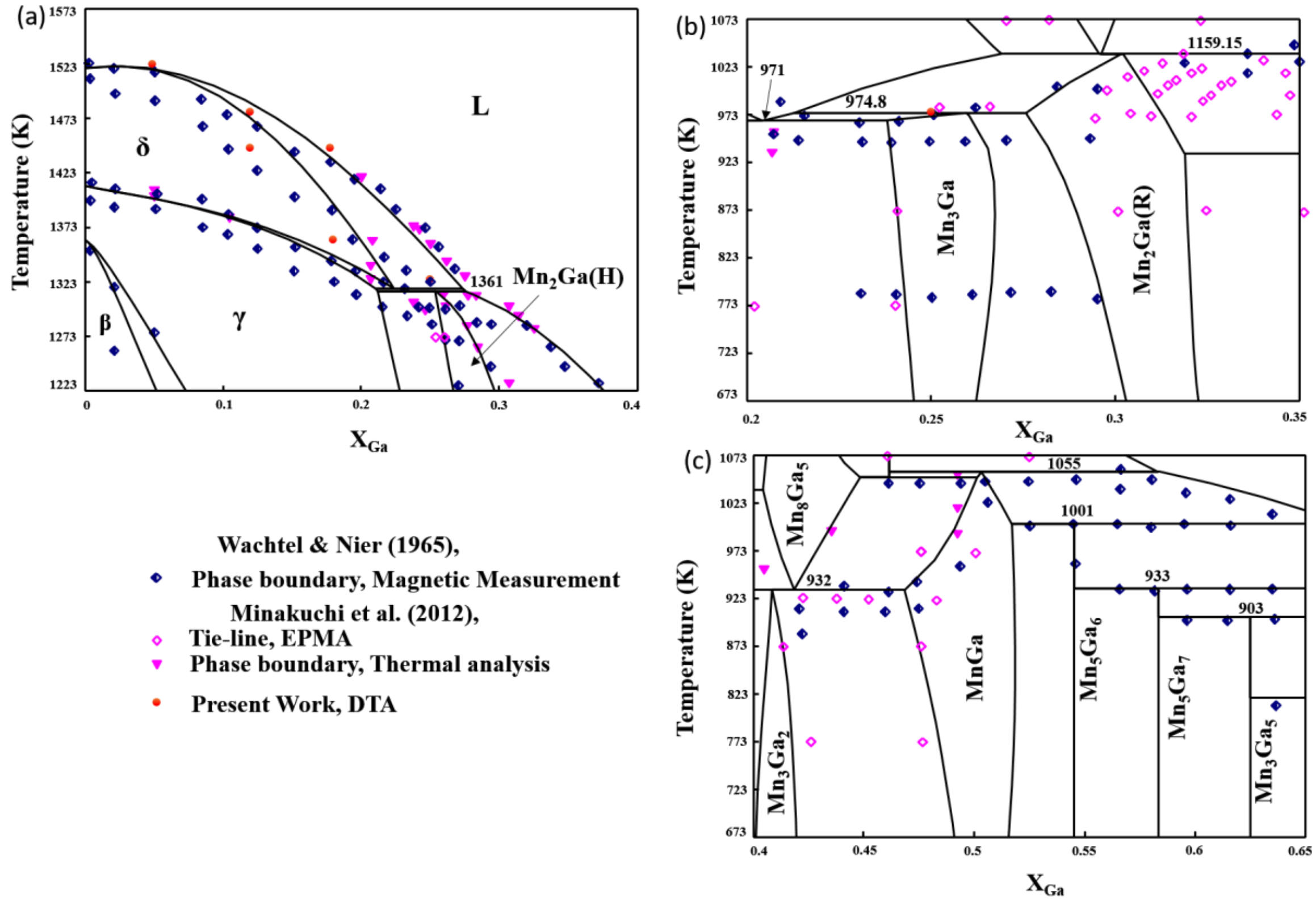

Fig. 4 (a), (b) and (c) shows the high temperature phase equilibria for Mn-rich side of the calculated phase diagram with present thermodynamic assessment and DTA results along with previous studies [51,59].

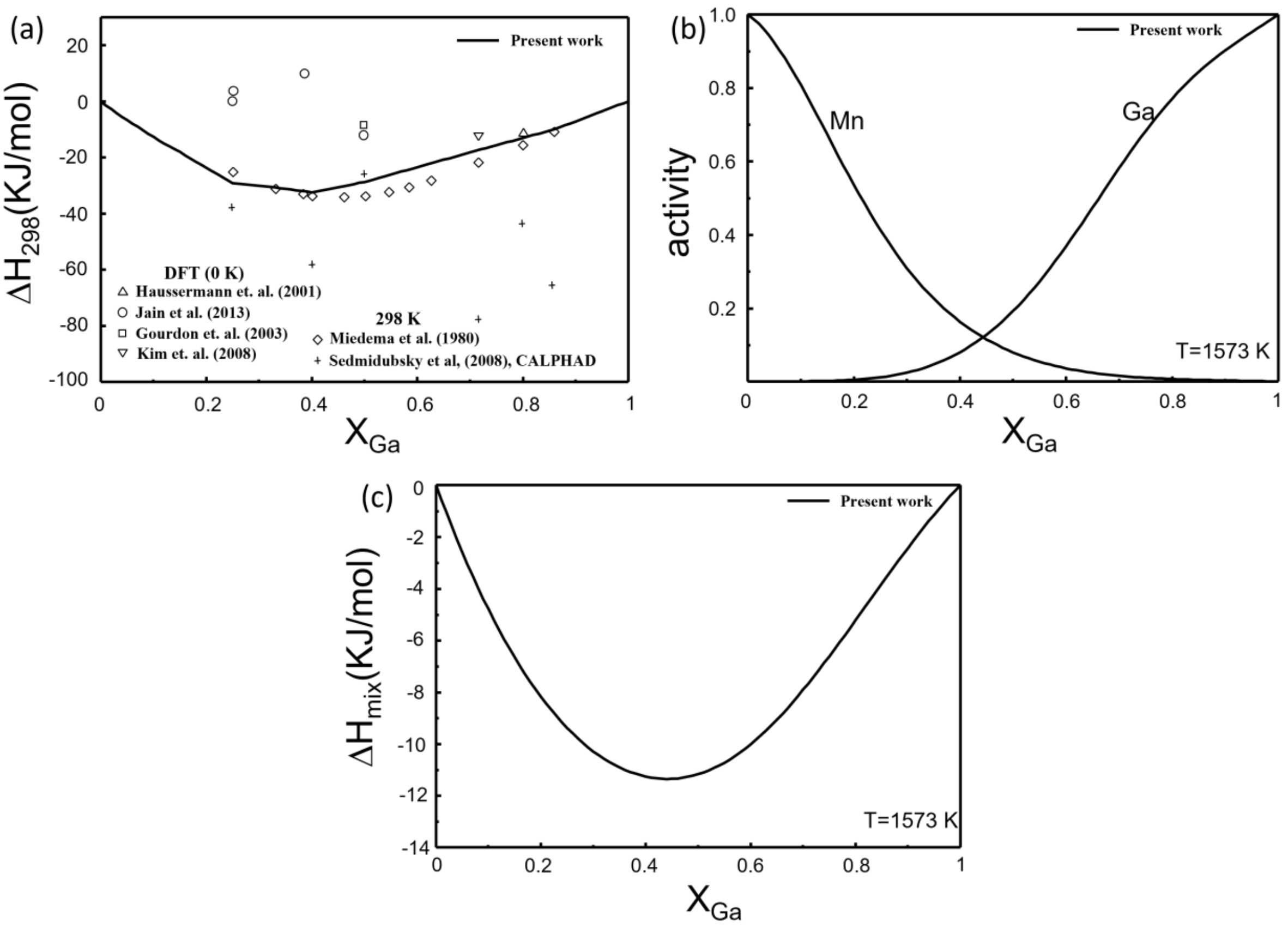

Fig. 5 (a) Calculated enthalpy of formation ($\Delta H_{298}$) at 298 K with available experimental data [64–66,68,69,87], (b) calculated activity of Mn and Ga in liquid at 1573 K with Ga (l) and Mn (l) as reference state and (c) calculated mixing enthalpy ($\Delta H_{mix}$) of liquid.

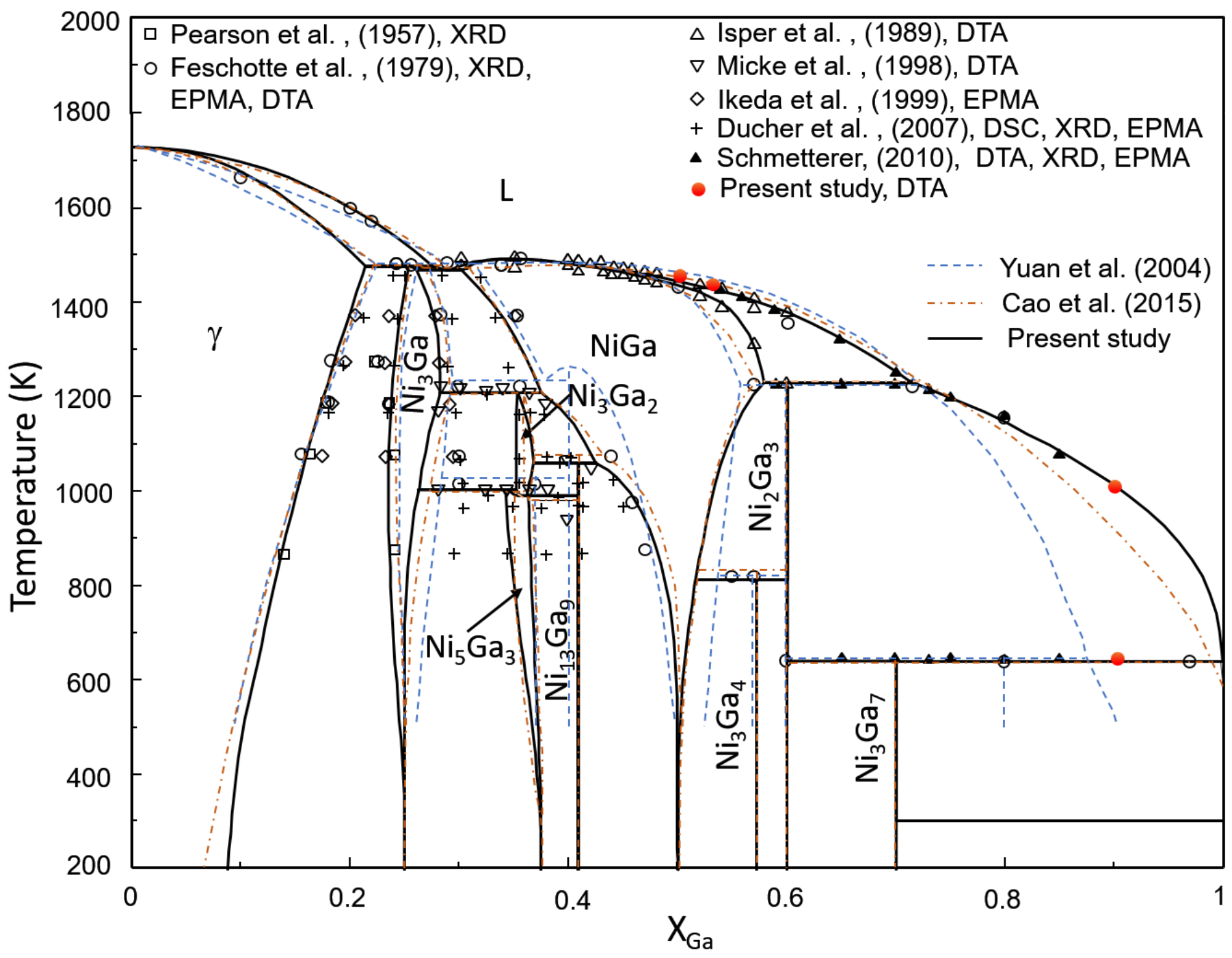

Fig. 6 Calculated phase diagram of Ni-Ga system along with experimental data [28,29,75,76,88].

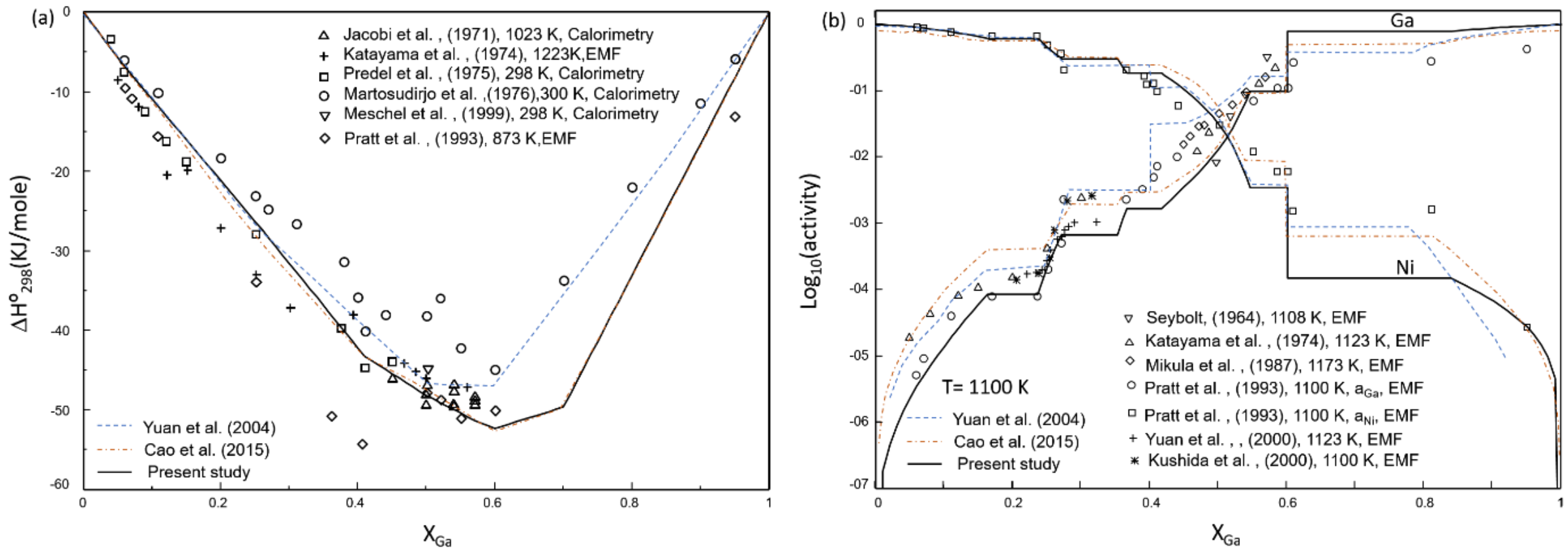

Fig. 7 Calculated (a) enthalpy of formation of Ni-Ga alloys with experimental data [77–81,84] and (b) $\log_{10}$(activity) of the Ni-Ga system at 1100 K with Ni(s) and Ga(l) as reference along with experimental data [80–83,85,86].

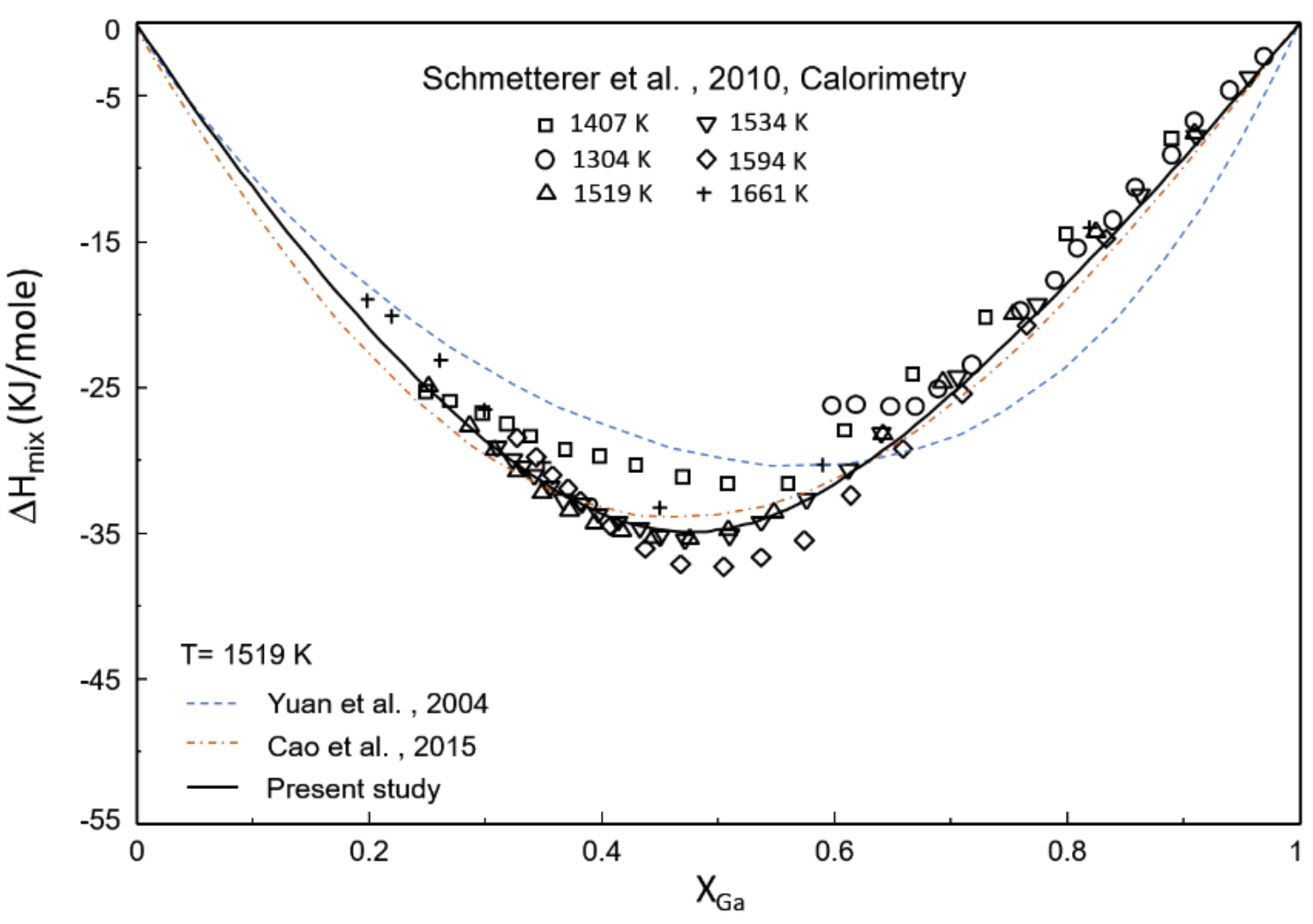

Fig. 8 Calculated mixing enthalpy of liquid at 1519 K compared with already reported calculated and experimental data [8, 10, 11].

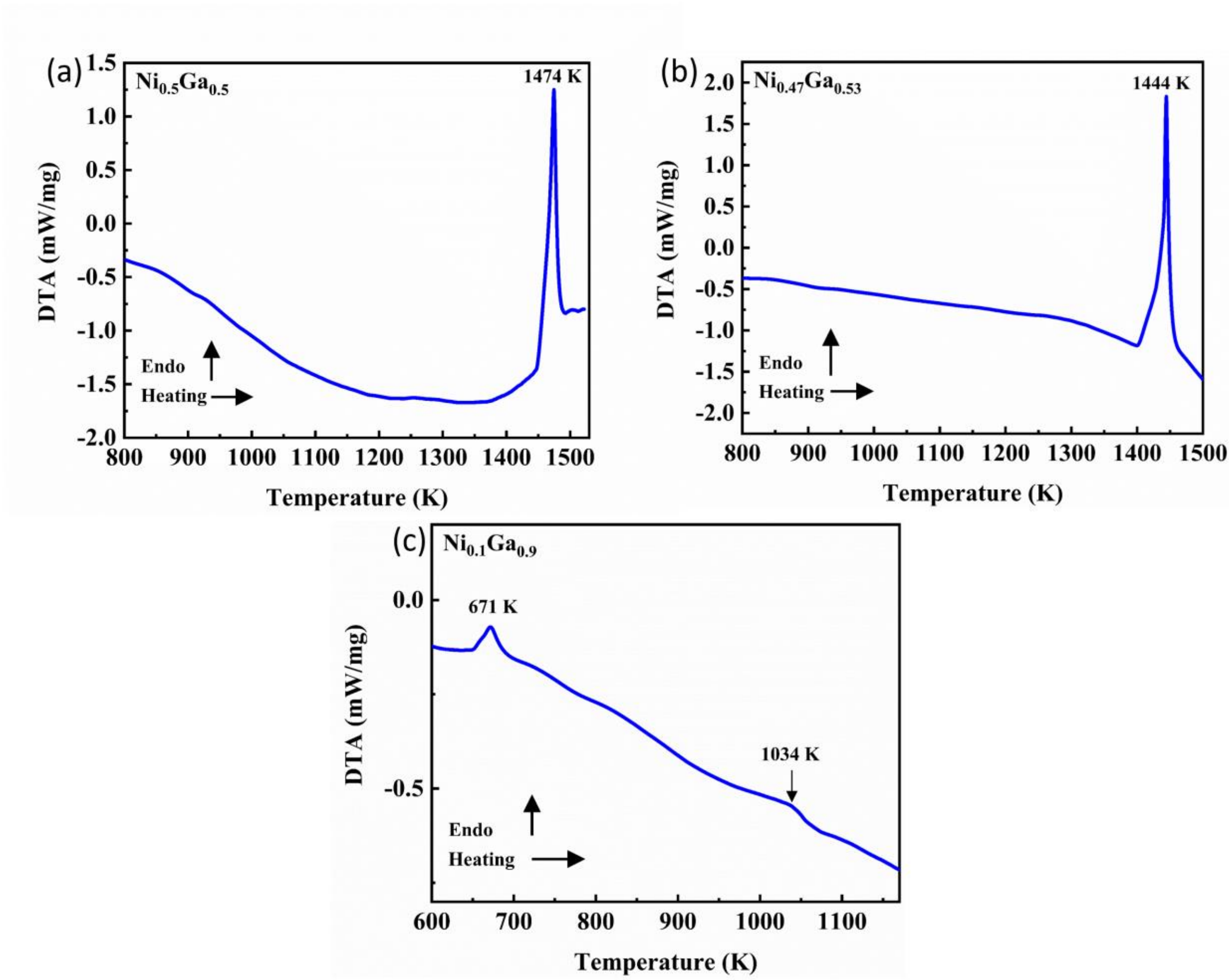

Fig. 9 DTA curve of (a) $Ni_{0.47}Ga_{0.53}$, (b) $Ni_{0.5}Ga_{0.5}$, and (c) $Ni_{0.1}Ga_{0.9}$ alloys.

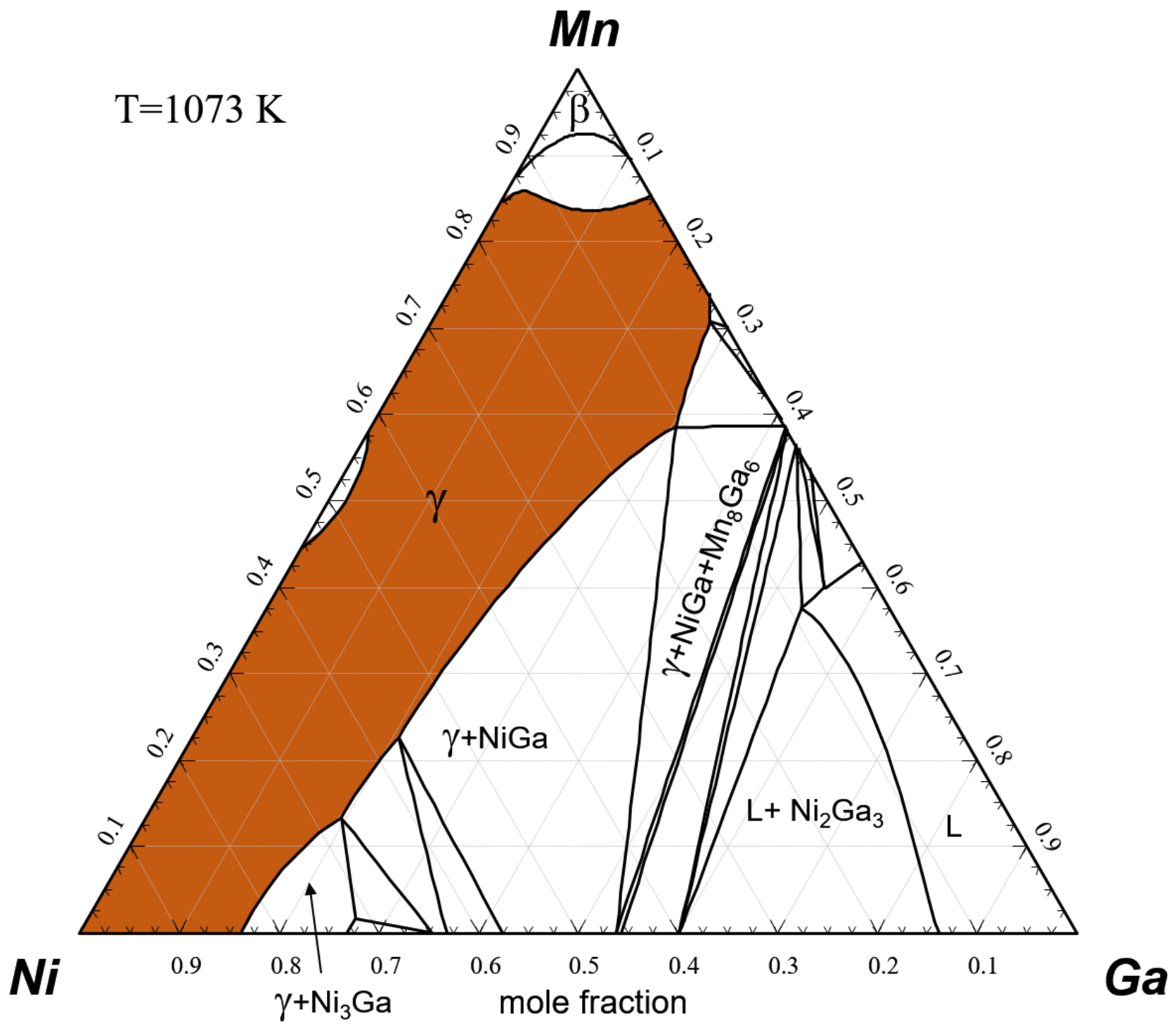

Fig. 10 Calculated isothermal section of Ni-Mn-Ga ternary system at 1073 K.

Table 1: Crystallographic information of the intermetallic compounds of Mn-Ga binary system and their sublattice model used in present CALPHAD modelling.

| Phase name (Wyckoff sites, Space Group) | X | Y | Z | Model (Present Study) | References |
|---|---|---|---|---|---|
| **1- $Mn_3Ga$** | | | | **$(Mn, Ga)_3$ (Mn, Ga)** | |
| a) Hexagonal* *($P6_3/mmc$)* | | | | | [49] |
| 2c | 0.66 | 0.33 | 0.75 | | |
| 6h | 0.15 | 0.31 | 0.75 | | |
| b) Cubic *(Fm3m)* | | | | | [46] |
| 4a | 0 | 0 | 0 | | |
| 4b | 0.50 | 0 | 0 | | |
| 8c | 0.25 | 0.25 | 0.75 | | |
| c) Tetragonal *(I4/mmm)* | | | | | [49] |
| 2a | 0 | 0 | 0 | | |
| 2b | 0.50 | 0.50 | 0 | | |
| 4d | 0 | 0.50 | 0.25 | | |
| **2- $Mn_8Ga_5$** | | | | | |
| a) Tetragonal* ($L1_0$) | | | | **$(Mn, Ga)_4$ $(Mn, Ga)_3$ $(Mn)_4$ $(Mn, Ga)_2$** | [56] |
| b) Cubic *(I43m)* | | | | | [57] |
| 8c | 0.89 | 0.10 | 0.10 | | |
| 8c | 0.67 | 0.32 | 0.32 | | |
| 12e | 0.64 | 0 | 0 | | |
| 24g | 0.18 | 0.81 | 0.53 | | |
| **3- MnGa** | | | | | |
| a) AuCu* *(P4/mmm)* | | | | **(Mn, Ga) (Mn, Ga)** | [62] |
| 1a | 0 | 0 | 0 | | |
| 1d | 0.50 | 0.50 | 0.50 | | |
| b) Hexagonal *(R3m)* | | | | | [47] |
| c) Monoclinic *(mS276)* C12/c1 | | | | | [51] |
| **4- $Mn_2Ga$ (R)** | | | | | |
| AuCu* *(P4/mmm)* | | | | **$(Mn, Ga)_2$ (Mn, Ga)** | [55] |
| 1a | 0 | 0 | 0 | | |
| 1d | 0.50 | 0.50 | 0.50 | | |
| **5- $Mn_2Ga$ (H)** | | | | | |
| Mg* *(P6$_3$/mmc)* | | | | **(Mn, Ga)$_2$ (Mn, Ga)** | [55] |
| **6- Mn$_3$Ga$_2$** | | | | | |
| AuCu* *(P4/mmm)* | | | | **(Mn, Ga)$_3$ (Mn, Ga)$_2$** | [51] |
| 1a | 0 | 0 | 0 | | |
| 1d | 0.50 | 0.50 | 0.50 | | |
| | | | | | |

** Crystal structure considered in present thermodynamic assessment of Mn-Ga binary system*

Table 2: Thermodynamically optimized model parameters for liquid and solid solution phases in Mn-Ga and Ni-Ga phase diagrams.

| | |
|---|---|
| Liquid<br>(Ni, Mn, Ga) (Va) | $Z^{Ni}_{NiNi} = Z^{Ga}_{GaGa} = Z^{Mn}_{MnMn}$=6,<br>$\Delta g_{MnGa}$ = -10700-5.25T + (-2850-3.80T) $X_{MnMn}$ + (-300) $X_{GaGa}$<br>+ (-5100) $X^2_{MnMn}$ + (4000) $X^2_{GaGa}$<br>$\Delta g_{NiGa} = -29800 + 2.28T + (-10000 - 4.10T)X_{NiNi} +$<br>$(-1800 + 3.26T)X_{GaGa}$<br>$\Delta g_{MnNi}$ = -11650 + (-1200+0.1T) $X_{MnMn}$ + (-850) $X_{NiNi}$ |
| β<br>(Mn, Ga) (Va) | $G_{Mn:Va} = {}^{\circ}G^{Cubic}_{Mn}$<br>$G_{Mn:Va} = {}^{\circ}G^{Cubic}_{Ga}$<br>$^0L_{Mn, Ga: Va}$ = -89000-7T<br>$^1L_{Mn, Ga: Va}$ = -99000 |
| γ<br>(Mn, Ni, Ga) (Va) | $G_{Mn:Va} = {}^{\circ}G^{FCC}_{Mn}$<br>$G_{Ga:Va} = {}^{\circ}G^{FCC}_{Ga}$<br>$G_{Ni:Va} = {}^{\circ}G^{FCC}_{Ni}$<br>$^0L_{Mn, Ga: Va}$ = -55000<br>$^1L_{Mn, Ga: Va}$ = -68000<br>$^2L_{Mn, Ga: Va}$ = -1000+12T<br>$^0L_{Ni, Ga; Va}$ = -125726 + 14T |
| α<br>(Mn, Ga) (Va) | $G_{Mn:Va} = {}^{\circ}G^{CBCC}_{Mn}$<br>$G_{Ga:Va} = {}^{\circ}G^{CBCC}_{Ga}$<br>$^0L_{Mn, Ga: Va}$ = -102000-28T<br>$^1L_{Mn, Ga: Va}$ = -45000 |
| δ | $G_{Mn:Va} = {}^{\circ}G^{BCC}_{Mn}$ |
| (Mn, Ga) (Va) | $G_{Mn:Va} = {}^{\circ}G_{Ga}^{BCC}$<br>${}^{0}L_{Mn, Ga: Va}$ = -33850-40T-0.01173$T^2$<br>${}^{1}L_{Mn, Ga: Va}$ = -20000 |
| $Mn_3Ga$<br>$(Mn, Ga)_3$ (Mn, Ga) | $G_{Mn: Mn} = 4{}^{\circ}G_{Mn}^{FCC}$ -2T<br>$G_{Mn: Ga} = {}^{\circ}G_{Ga}^{FCC} + 3{}^{\circ}G_{Mn}^{FCC}$ -134612 -52.70T<br>$G_{Ga: Mn} = 3{}^{\circ}G_{Ga}^{FCC} + {}^{\circ}G_{Mn}^{FCC}$-1000T<br>$G_{Ga: Ga} = 4{}^{\circ}G_{Ga}^{FCC}$<br>${}^{0}L_{Mn, Ga: Ga}$ = -38400<br>${}^{1}L_{Mn, Ga: Ga}$ = -95000<br>${}^{0}L_{Mn; Mn, Ga}$ = -20000<br>${}^{\circ}T_{C,Mn:Ga}^{Mn3Ga}$ = 470 K<br>${}^{\circ}\beta_{Mn:Ga}^{Mn3Ga}$ = 3.1 |
| $Mn_2Ga$(R)<br>$(Mn, Ga)_2$ (Mn, Ga) | $G_{Mn: Ga} = 2{}^{\circ}G_{Mn}^{FCC} + {}^{\circ}G_{Ga}^{FCC}$ -105425 -35.70T<br>$G_{Mn: Mn} = 3{}^{\circ}G_{Mn}^{FCC}$ + 100<br>$G_{Ga: Mn} = 2{}^{\circ}G_{Ga}^{FCC} + {}^{\circ}G_{Mn}^{FCC}$<br>$G_{Ga: Ga} = 3{}^{\circ}G_{Ga}^{FCC}$<br>${}^{0}L_{Mn; Mn, Ga}$ = -73000 + 37T<br>${}^{0}L_{Ga; Mn, Ga}$ = -30000<br>${}^{\circ}T_{C,Mn:Ga}^{Mn2Ga}$ = 690 K<br>${}^{\circ}\beta_{Mn:Ga}^{Mn2Ga}$ = 1.5 |
| $Mn_2Ga$ (H)<br>$(Mn, Ga)_2$ (Mn, Ga) | $G_{Mn: Ga} = 2{}^{\circ}G_{Mn}^{FCC} + {}^{\circ}G_{Ga}^{FCC}$ -46400 + 22.04T<br>$G_{Mn: Mn} = 3{}^{\circ}G_{Mn}^{FCC}$ + 100<br>$G_{Ga: Mn} = 2{}^{\circ}G_{Ga}^{FCC} + {}^{\circ}G_{Mn}^{FCC}$ + 100<br>$G_{Ga: Ga} = 3{}^{\circ}G_{Ga}^{FCC}$ |
| | $^{0}L_{Mn; Ga, Mn}$ = -23200 |
| MnGa<br><br>(Mn, Ga) (Mn, Ga) | $G_{Mn: Mn} = 2°G_{Mn}^{FCC} + 10000$<br><br>$G_{Mn: Ga} = °G_{Ga}^{FCC} + °G_{Mn}^{FCC} -65500 - 15.70T$<br><br>$G_{Ga: Mn} = °G_{Ga}^{FCC} + °G_{Mn}^{FCC} - 20000$<br><br>$G_{Ga: Ga} = 2°G_{Ga}^{FCC} + 1400 - 16.60T$<br><br>$^{0}L_{Mn, Ga: Ga}$ = -1100 - 17T<br><br>$^{0}L_{Mn; Mn, Ga}$ = -88000 + 59T<br><br>$^{0}L_{Ga; Mn, Ga}$ = -10000<br><br>$^{1}L_{Mn; Mn, Ga}$ = 1000 |
| $Mn_8Ga_5$<br><br>$(Mn, Ga)_4 (Mn, Ga)_3$<br><br>$(Mn)_4 (Mn, Ga)_2$ | $G_{Mn: Mn: Mn; Mn} = 13°G_{Mn}^{BCC} + 10T$<br><br>$G_{Mn: Ga: Mn: Mn} = 3°G_{Ga}^{BCC} + 10°G_{Mn}^{BCC}$<br><br>$G_{Mn: Mn: Mn: Ga} = 2°G_{Ga}^{BCC} + 11°G_{Mn}^{BCC}$<br><br>$G_{Mn: Ga: Mn: Ga} = 5°G_{Ga}^{BCC} + 8°G_{Mn}^{BCC} -214000- 162.2T$<br><br>$G_{Ga: Mn: Mn: Mn} = 4°G_{Ga}^{BCC} + 9°G_{Mn}^{BCC}$<br><br>$G_{Ga: Ga: Mn: Mn} = 7°G_{Ga}^{BCC} + 6°G_{Mn}^{BCC} + 10000$<br><br>$G_{Ga: Mn: Mn: Ga} = 6°G_{Ga}^{BCC} + 7°G_{Mn}^{BCC}$<br><br>$G_{Ga: Ga: Mn: Ga} = 9°G_{Ga}^{BCC} + 4°G_{Mn}^{BCC}$<br><br>$^{0}L_{Mn, Ga; Ga; Mn; Mn}$ = -235183 - 251T<br><br>$^{0}L_{Ga; Ga; Mn; Mn, Ga}$ = -235183- 350T<br><br>$°T_{C,Mn:Ga}^{Mn8Ga5} = 210$ K<br><br>$°\beta_{Mn:Ga}^{Mn8Ga5} = 1.98$ |
| $Mn_3Ga_2$<br><br>$(Mn, Ga)_3 (Mn, Ga)_2$ | $G_{Mn:Mn} = 5°G_{Mn}^{FCC} + 5000$<br><br>$G_{Mn:Ga} = 2°G_{Ga}^{FCC} + 3°G_{Mn}^{FCC} -183000 - 64.30T$<br><br>$G_{Ga:Mn} = 3°G_{Ga}^{FCC} + 2°G_{Mn}^{FCC} + 14000$ |
| | $G_{Ga:Ga} = 5°G_{Ga}^{FCC} + 19600$<br>$^{0}L_{Mn: Mn, Ga} = -45000$<br>$^{0}L_{Mn, Ga: Ga} = -92000$<br>$°T_{C,Mn:Ga}^{Mn3Ga2} = 630$ K<br>$°\beta_{Mn:Ga}^{Mn3Ga2} = 1.25$ |
| $Ni_3Ga$<br>$(Ni, Ga)_3 (Ni, Ga)$ | $G_{Ni;Va} = °G_{Ni}^{FCC} + 22000$<br>$G_{Ga;Va} = °G_{Ga}^{FCC}$<br>$G_{Ni;Ga} = 3°G_{Ni}^{FCC} + °G_{Ga}^{FCC} - 112306 - 5.3T$<br>$G_{Ga;Ni} = °G_{Ni}^{FCC} + 3°G_{Ga}^{FCC} - 112306 + 5.3T$<br>$^{0}L_{Ni,Ga;Ga} = {}^{0}L_{Ni,Ga;Ni} = -222312 + 13.12T$<br>$^{1}L_{Ni,Ga;Ga} = -15000 + 5T$<br>$^{1}L_{Ni;Ni,Ga} = {}^{1}L_{Ga;Ni,Ga} = 4020 + 5T$<br>$^{0}L_{Ni;Ni,Ga} = -4160 - 6T$ |
| NiGa<br>(Ga, Ni) (Va, Ni) | $G_{Ni;Va} = °G_{Ni}^{BCC}$<br>$G_{Ga;Va} = °G_{Ga}^{BCC} + 33200 + 2T$<br>$G_{Ni;Ni} = 2°G_{Ni}^{BCC} + °G_{Ga}^{BCC} + 34900 - 7.60T$<br>$G_{Ga;Ni} = °G_{Ni}^{BCC} + °G_{Ga}^{BCC} - 111360.5 + 27.25T$<br>$^{0}L_{Ga,Ni;Ni} = -56000 - 15.40T$<br>$^{0}L_{Ga,Ni;Va} = -64500 - 8.42T$<br>$^{0}L_{Ga; Va, Ni} = -64500 - 4.72T$<br>$^{1}L_{Ga,Ni;Ni} = 1800 + 3T$<br>$^{1}L_{Ga;Va,Ni} = 8500 - 10T$ |
| $Ni_3Ga_2$ | $G_{Ga;Ni;Ni} = 2°G_{Ni}^{SER} + °G_{Ga}^{SER} - 102478 + 0.67T$<br>$G_{Ga;Ni;Va} = °G_{Ga}^{SER} + °G_{Ni}^{SER} - 78900 + 1.50T$ |
| (Ga, Ni) (Ni, Va) (Va, Ni) | $G_{Ga;Va;Ni} = {}^{\circ}G_{Ga}^{SER} + {}^{\circ}G_{Ni}^{SER} - 13500 + 15T$<br>$G_{Ga;Va;Va} = {}^{\circ}G_{Ga}^{SER} - 22500 + 22T$<br>$G_{Ni;Ni;Ni} = 3{}^{\circ}G_{Ni}^{SER} + 10000 + 10T$<br>$G_{Ni;Va;Ni} = G_{Ni;Ni;Va} = 2{}^{\circ}G_{Ni}^{SER} + 10000 + 10T$<br>$G_{Ni;Va;Va} = {}^{\circ}G_{Ni}^{SER} + 10000 + 10T$<br>${}^{0}L_{Ga;Ni,Va;Ni} = {}^{0}L_{Ni;Ni,Va;Va} = -13500 - 9T$<br>${}^{1}L_{Ga;Ni,Va;Ni} = {}^{1}L_{Ga;Ni,Va;Va} = 2700 + 3T$<br>${}^{0}L_{Ga;Ni;Va,Ni} = {}^{0}L_{Ga;Va;Va,Ni} = -29700 + 27.75T$<br>${}^{1}L_{Ga;Ni;Va,Ni} = {}^{1}L_{Ga;Va;Va,Ni} = -1500$ |
| $Ni_5Ga_3$<br>$(Ni)_5\,(Ni, Ga)_3$ | $G_{Ni;Ni} = 8{}^{\circ}G_{Ni}^{SER} + 20000$<br>$G_{Ni;Ga} = 5{}^{\circ}G_{Ni}^{SER} + 3{}^{\circ}G_{Ga}^{SER} - 325644 + 28.20T$<br>${}^{0}L_{Ni;Ni,Ga} = 12000 - 59T$ |
| All are in $(J/(mol.K)^{-1})$. | |

Table 3: Model parameters for intermetallic compounds of Mn-Ga and Ni-Ga phase diagrams.

| Compound | $H_{298}$ (kJ/mole) | $S_{298}$ (J/(mol.K)) | Cp (T) | T (K) |
|---|---|---|---|---|
| $MnGa_4$ | -65.35 | 220.43 | $-29.18+ 0.17T-8929929.43T^{-2}$ $+62762.29^{-1}$ | 298-500 |
| | | | $137.35+ 0.014T-157054.12T^{-2}$ | 500-980 |
| | | | $147.74+ 0.004T-343245.57T^{-2}$ | 980-1421 |
| | | | 159.53 | 1421-1422 |
| $MnGa_6$ | -71.55 | 313.35 | $-55.70+0.24T-13316367.09T^{-2}+$ $94143.49T^{-1}$ | 298.15-500 |
| | | | $194.10+0.01T-157054.12T^{-2}$ | 500-980 |
| | | | $204.49+ 0.004T-343245.57T^{-2}$ | 980-1421 |
| | | | 216.23 | 1421-1422 |
| $Mn_2Ga_5$ | -122.40 | 295.18 | $-18.59+0.22T-11280202.38T^{-2}$ $+78452.86T^{-1}$ | 298.15-500 |
| | | | $189.58+0.03T-314108.23T^{-2}$ | 500-980 |
| | | | $210.35+0.008T-686491.15T^{-2}$ | 980-1421 |
| $Mn_3Ga_5$ | -177.5 | 318.35 | $5.26+ 0.23T- 11437256.50T^{-2}+$ $78452.86T^{-1}$ | 298.15-500 |
| | | | $213.43+ 0.04T- 471162.35T^{-2}$ | 500-980 |
| | | | $244.59+ 0.01T$ $- 1029736.72T^{-2}$ | 980-1421 |
| | | | 279.95 | 1421-1422 |
| $Mn_5Ga_7$ | -292.70 | 461.10 | 26.44+ 0.34T- 16137802.39$T^{-2}$+ 109834$T^{-1}$ | 298.15-500 |
| | | | 317.88+ 0.07T- 785270.58$T^{-2}$ | 500-980 |
| | | | 369.82+ 0.02T- 1716227.86$T^{-2}$ | 980-1421 |
| | | | 428.75 | 1421-1422 |
| $Mn_5Ga_6$ | -290.12 | 408.06 | 39.70+ 0.30T- 13944583.56$T^{-2}$+ 94143.43$T^{-1}$ | 298.15-500 |
| | | | 289.50+ 0.07T- 785270.58$T^{-2}$ | 500-980 |
| | | | 341.45+ 0.02T- 1716227.86 | 980-1412 |
| | | | 338.08+ 0.04T | 1412-1900 |
| | | | 400.37 | 1900-1901 |
| $Mn_7Ga_6$ | -219.10 | 600.00 | 87.40+ 0.33T-14258691.79$T^{-2}$+ 94143.43$T^{-1}$ | 298.15-500 |
| | | | 337.20+ 0.10T-1099378.82$T^{-2}$ | 500-980 |
| | | | 409.92+ 0.03T-2402719.01$T^{-2}$ | 980-1421 |
| | | | 492.42 | 1421-1422 |
| $Ga_3Ni_2$ | -262.47 | 139.74 | 368.88-1.34T-2639724$T^{-2}$ +0.002$T^2$ | 298.15-302.92 |
| | | | 122.40 + 0.018T+709992 $T^{-2}$ +7.23E-7$T^2$ +4.44E25$T^{-10}$ | 302.90-1728 |
| $Ga_4Ni_3$ | -358.70 | 202.44 | 499.20 -1.79 T -3519632 $T^{-2}$+ 0.003 $T^2$ | 298.15-302.92 |
| | | | 170.56+ 0.03T + 946656 $T^{-2}$+ 9.64E-7 $T^2$ -5.92E25 $T^{-10}$ | 302.92-1728 |
| $Ga_7Ni_3$ | -496.16 | 176.04 | $823.89-3.15\ T -6159356\ T^{-2} +0.005\ T^{2}$ | 298.15-302.92 |
| | | | $248.77+0.03T +1656648\ T^{-2} +1.69E\text{-}6\ T^{2}-1.04E26\ T^{-10}$ | 302.92-1728 |
| | | | $311.78-0.002\ T +1656648\ T^{-2} +1.69E\text{-}6\ T^{2}-3.04E33\ T^{-10}$ | 1728-3000 |
| $Ga_9Ni_{13}$ | -949.59 | 689.85 | $1261.31-3.96\ T -7919172\ T^{-2} +0.006\ T^{2}$ | 298.15-302.92 |
| | | | $521.87+0.12\ T +2129976\ T^{-2} +2.17E\text{-}6\ T^{2}-1.33E26\ T^{-10}$ | 302.92-1728 |
| | | | $794.92-0.003\ T +2129976\ T^{-2} +2.17E\text{-}6\ T^{2}-1.32E34\ T^{-10}$ | 1728-3000 |
| | | | $794.70-0.003\ T +2129976\ T^{-2} +2.17E\text{-}6\ T^{2}-1.33E26\ T^{-10}$ | 3000-4000 |